# Fast whole-brain imaging of seizures in zebrafish larvae by two-photon light-sheet microscopy

Giuseppe de Vito[1,2,†], Lapo Turrini[2,3,†], Caroline Müllenbroich[2,4,5], Pietro Ricci[2], Giuseppe Sancataldo[2,3], Giacomo Mazzamuto[2,5], Natascia Tiso[6], Leonardo Sacconi[2,5], Duccio Fanelli[3], Ludovico Silvestri[2,3,5], Francesco Vanzi[2,7], Francesco Saverio Pavone[2,3,5,*]

[1]*University of Florence, Department of Neuroscience, Psychology, Drug Research and Child Health, Viale Pieraccini 6, Florence, Italy, 50139*
[2]*European Laboratory for Non-Linear Spectroscopy, Via Nello Carrara 1, Sesto Fiorentino, Italy, 50019*
[3]*University of Florence, Department of Physics and Astronomy, Via Sansone 1, Sesto Fiorentino, Italy, 50019*
[4]*School of Physics and Astronomy, Kelvin Building, University of Glasgow, G12 8QQ, Glasgow, UK*
[5]*National Institute of Optics, National Research Council, Via Nello Carrara 1, Sesto Fiorentino, Italy, 50019*
[6]*University of Padova, Department of Biology, Via U. Bassi 58/B, Padova, Italy, 35131*
[7]*University of Florence, Department of Biology, Via Madonna del Piano 6, Sesto Fiorentino, Italy, 50019*
[†]*Co-first authors with equal contribution*
[*]*francesco.pavone@unifi.it*

**Abstract:** Light-sheet fluorescence microscopy (LSFM) enables real-time whole-brain functional imaging in zebrafish larvae. Conventional one photon LSFM can however induce undesirable visual stimulation due to the use of visible excitation light. The use of two-photon (2P) excitation, employing near-infrared invisible light, provides unbiased investigation of neuronal circuit dynamics. However, due to the low efficiency of the 2P absorption process, the imaging speed of this technique is typically limited by the signal-to-noise-ratio. Here, we describe a 2P LSFM setup designed for non-invasive imaging that enables quintuplicating state-of-the-art volumetric acquisition rate of the larval zebrafish brain (5 Hz) while keeping low the laser intensity on the specimen. We applied our system to the study of pharmacologically-induced acute seizures, characterizing the spatial-temporal dynamics of pathological activity and describing for the first time the appearance of caudo-rostral ictal waves (CRIWs).

## 1. Introduction

Recent scientific advances, both in microscopy technology [1–5] and in fluorescent sensors of neuronal activity [6,7], have led to a profound revolution in the field of brain functional imaging, making it pivotal for understanding brain functions. Deciphering the workings of a brain (and its alterations due to pathologies and disorders) requires real-time mapping of the activity of all neurons and in this scenario light-sheet fluorescence microscopy (LSFM) [8] plays a crucial role. Indeed, owing to its unique architecture, which couples optical sectioning ability with parallelization of the acquisition process within each frame, LSFM has allowed for the first time the high-speed volumetric imaging of a vertebrate brain in its entirety, employing the intrinsically transparent zebrafish larva as a sample [9]. In particular, this technique has enabled the simultaneous functional investigation of large populations of neurons throughout the entire zebrafish brain expressing fluorescent calcium indicators, leading to novel insights into circuit dynamics [9–14].

Conventional LSFM employs continuous-wave lasers as sources for one-photon excitation and, apart from the use of peculiar illumination tailoring aimed at excluding eye exposure [15], this usually leads to detrimental visual stimulation during imaging. At best, the use of a visible light sheet may induce undesired activation of retino-recipient neurons throughout the brain, making it hard to disentangle this activity from those under investigation [16,17]. At worst, it may preclude neuronal activity measurement in more delicate experimental conditions, i.e. more susceptible to photostimulation, such as target-driven fictive navigation, sleep/wake rhythm or specific pathological states. The use of two-photon (2P) excitation [18] instead, employing near infrared pulsed light, which is invisible to the majority of vertebrates (including zebrafish [19]), allows for a potential expansion of the experimental panorama amenable to high-speed imaging. However, owing to the low probability of the 2P absorption process, 2P LSFM [20–22] is prone to low signal-to-noise ratio (SNR) issues, which remarkably limit the actual volumetric acquisition speed [23]. For this reason, thus far, 2P LSFM has been proficiently applied to perform structural studies [24], single plane functional imaging [25,26] and functional studies at low volumetric rate [16,17,27,28]. To date, no high-speed volumetric functional imaging has been performed by means of 2P LSFM. Here, we describe a 2P LSFM setup achieving fast whole-brain imaging (5 volumes per second) in zebrafish larvae [29,30] through SNR boosting, while keeping low the incident power on the sample, a crucial aspect for *in-vivo* imaging [23,31].

We applied the system to investigate the neuronal dynamics occurring in the larval zebrafish brain during epileptic seizures. Seizures are recurrent and unprovoked episodes of paroxysmal neuronal activity typically occurring as a consequence of a loss of balance between excitatory and inhibitory synaptic activity and they represent the hallmark of epilepsy. We induced seizures in 4 days post fertilization (dpf) larvae using pentylenetetrazole (PTZ), a molecule employed to model seizure-like behavior in zebrafish larvae [32,33] thus modelling several kinds of human epilepsies in this animal [34].

We mapped in real-time, on a brain-wide scale and with cellular resolution, the onset and propagation of acute seizures, at the same time avoiding detrimental photostimulation on a highly susceptible system such as an epileptic brain. In this work, we show that the presented 2P LSFM design enables high spatio-temporal resolution mapping of seizure dynamics, allowing the emergence of peculiar postero-anterior propagation patterns to be revealed for the first time.

## 2. Experimental techniques and materials

*2.1 2P-LSFM setup*

The employed 2P LSFM setup is depicted in Fig. 1. A tunable Ti-Sa pulsed laser (Chameleon Ultra II, Coherent) is used to generate the excitation light at 930 nm. The laser is coupled with a pulse compressor unit (PreComp, Coherent) to pre-compensate for the group delay dispersion (induced by the optics and by the long path travelled by the focused light after the excitation objectives inside the water-filled imaging chamber), thus increasing the SNR. At the exit of the pulse compressor, the laser beam is attenuated using a half-wave plate and a Glan-Thompson polarizer that dumps part of the beam.

The beam then passes through an electro-optical modulator (EOM, 84502050006, Qioptiq) that switches the polarization state of the light between two orthogonal states with a frequency of 100 kHz. After the EOM, a quarter-wave plate is used to pre-compensate for the polarization distortions and a half-wave plate is employed to rotate the couple of light polarization planes in order to be parallel or perpendicular with respect to the optical table surface.

The light is then routed to a hybrid pair of galvanometric mirrors through a periscope and a pair of steering mirrors. One of the galvanometric mirrors is a fast resonant mirror (CRS-8 kHz, Cambridge Technology) and it is used to scan the beam along the fronto-caudal direction of the larva with a frequency of 8 kHz to digitally generate the light sheet. The other galvanometric mirror is a closed-loop mirror (6215H, Cambridge Technology) and it is used to scan the light-sheet along the dorso-ventral direction of the larva with a frequency of 5 Hz.

Two excitation dry objectives are placed at the lateral sides of the larva, outside the sample chamber. The scanned beam is routed to one of the two excitation objectives (XLFLUOR4X/340/0,28, Olympus) by means of a scan lens (50 mm focal length), a tube lens (75 mm focal length) and a pair of relay lenses (250 mm and 200 mm focal lengths). The lens series after the galvos magnifies the beam diameter 1.2 times, therefore the objective pupils are underfilled. When the incoming light is polarized perpendicularly to the table surface by the EOM, it is diverted by a polarizing beam splitter placed between the tube lens and the first relay lens to illuminate the second excitation arm. Immediately after this beam splitter, a half-wave plate is used to rotate the light polarization plane so that the light from both the excitation objectives is polarized parallel to the table surface. Finally, a pair of relay lenses identical to the one on the first excitation arm is used to route the beam to the second excitation objective.

The excitation light is focused inside the custom-made sample chamber filled with fish water. This chamber is walled by thin (0.17 mm) glass surfaces and is open on the top. It is thermostated (at 28.5 °C) using a dedicated custom closed-loop controller. The chamber is designed to hold a modified glass slide having a raised glass wedge at the center, on the top of which the larva is placed. In this way the focused light beams are not distorted by the presence of the adjacent glass slide surface.

The fine three-dimensional placement of the sample under the detection objective is performed by moving the sample chamber with three micro-positioning stages (M-521.DD1, M-511.DD1 and M-501.1DG, Physik Instrumente) while observing the detection-objective relative position with an infrared-sensible auxiliary camera (UI-1240SE-NIR-GL, iDS) placed frontally to the larva, outside the chamber.

A water-immersion high-numerical aperture (NA) detection objective (XLUMPLFLN20XW, Olympus, NA=1) placed on the dorsal side of the larva is used to collect fluorescence emission. For this objective we use a tube lens with a focal length of 300 mm. The collected light is then routed to the sCMOS camera (ORCA-Flash4.0 V3, Hamamatsu) by a relay system composed by an additional tube lens (focal lens: 200 mm) and a dry objective (UPLFLN10X2, Olympus, NA=0.3). The total magnification of the detection arm is 3×.

Immediately before the camera objective pupil we placed the filter wheel and an electrically-tunable lens (ETL). The ETL (EL-16-40-TC-VIS-5D-C, Optotune) is used to remotely scan the focal plane of the sample-side detection objective in synchrony with the light-sheet motion. To select GCaMP6s emission we used a green fluorescence filter (FF01-510/84-25 nm BrightLine® single-band bandpass filter, Semrock) placed in the filter wheel.

The two excitation objectives and the sample-side detection objective are mounted on objective scanners (PIFOC P-725.4CD, Physik Instrumente). Before each acquisition, the axial positions of the objectives are tuned in order to overlap the scanning range of the ETL with that of the closed-loop galvanometric mirror and to homogeneously illuminate the sample from the sides.

## 2.2 Acquisition control

The microscope is controlled by two workstations (Precision Tower 5810 and Precision Tower 7810, Dell), one of which is dedicated exclusively to the camera control, exploiting a frame-grabber device and a RAID-0 array of four fast SSD devices, while the other one is used to manage all the other microscope hardware, also using a data acquisition device (PCIe-6353, National Instruments). Image acquisition and instrument control are performed by dedicated custom software written in "G" (LabVIEW, National Instruments).

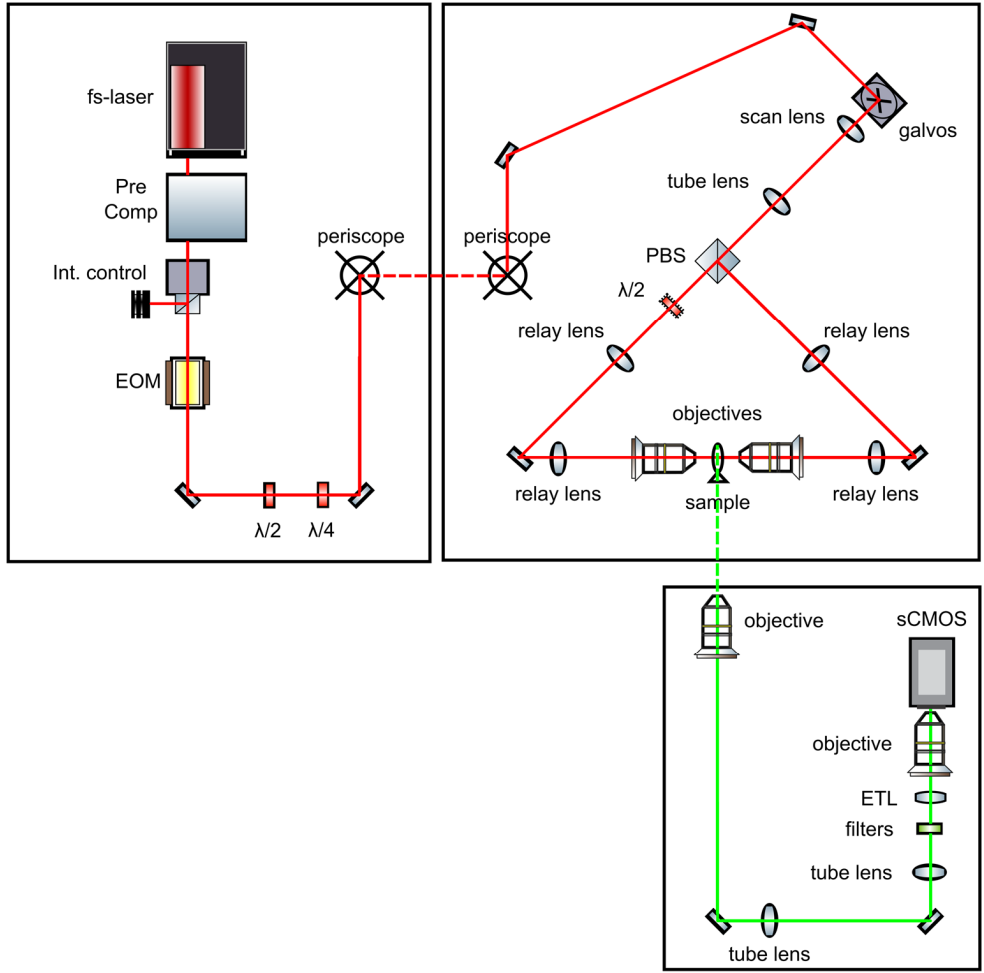

Fig. 1. Schematic of the novel custom-made 2P LSFM setup. fs-laser: femtosecond laser. Pre Comp: group delay dispersion precompensation unit. Int. control: power intensity control unit. EOM: electro-optic modulator. λ/2: half-wave plate. λ/4: quarter-wave plate. galvos: hybrid galvanometric mirrors assembly. PBS: polarizing beam splitter. ETL: electrically-tunable lens. Red lines: excitation beampath, green lines: detection beampath. Dashed segments indicate vertical paths.

*2.3 Optical system characterization*

Detection point spread function (PSF) measurements were performed in two ways. First, we visualized fluorescent beads (T7280, Invitrogen) with a diameter of 2.5 µm included in agarose gel, employing a blue LED source for excitation. With this method, we measured a transversal full width at half maximum (FWHM) of 6 µm and an axial FWHM of 10 µm. Then, we observed with transmitted light a grid made by horizontal and vertical lines spaced 10 µm (R1L3S3P, Thorlabs). We computed the transversal PSF size from the two-dimensional Fourier transform of this image and a size of 5.2 µm resulted from the modulation transfer function. This 13% discrepancy can be attributed to the size of the employed beads in the first method.

Excitation beam characterizations were performed by imaging fluorescence emission induced in a fluorescein solution. The transversal fluorescence profiles at each position along the beam axial dimension were fitted with Gaussian functions. From the fit parameters we obtained the maxima and the transversal FWHMs for each longitudinal position. Finally, we obtained the beam maximum fluorescence value and longitudinal FWHM by fitting the computed transversal maxima with a Gaussian function. With this method we found a transversal FWHM at the waist of 6 μm (Fig. S1a,b) and a longitudinal FWHM of 327 μm (Fig. S1c).

To test the advantages of employing the emi-field alternating illumination, we replaced the polarizing beam-splitter on the illumination path with a non-polarizing beam-splitter and we deactivated the EOM controller. Moreover, we rotated the half-wave plate after the beam-splitter in order to maintain horizontal polarization in both the excitation arms. The maximum fluorescence values in the two conditions were computed with the same procedure described above (Fig. S1c,d). The measure was performed with a cumulative excitation power of 100 mW (after the objectives). Before the measure we verified the cumulative fluorescence signal displaying the expected quadratic dependence on the excitation power up to 205 mW by fitting it with a parabolic function, finding a determination coefficient equal to 0.999.

### 2.4 Laser power

For all the imaging sessions, the cumulative power of the excitation beams at the exit of the two objectives was set at 200 mW ± 5% (at 100 mW ± 5% for each objective in time-average). These values were chosen after performing preliminary studies about phototoxicity. In brief, we continuously imaged 10 larvae for extended periods of time (more than 1 hour) and with excitation powers varying between 125 mW and 210 mW. For each larva we reported the time when photodamage appeared in the brain volume as areas of tonic activation of the calcium sensor (Fig. S2 and Table S1), typically located in the lateral (left or right) optic tectum. We observed that 60% of the larvae continuously exposed to uninterrupted imaging with illumination power ≥ 190 mW did not show photodamage signs in the first < 30 minutes.

We estimated the power loss induced by the passage through the glass-walled sides of the sample camera by measuring the laser power levels before and after the glass wall in air (without fish-water) and we found it to correspond to ~10%. Finally, employing the Lambert-Beer's law and considering the water absorption coefficient at 930 nm to be 0.128 cm$^{-1}$ (as reported by Kou and collaborators [35]), we calculated that the 25 mm-path travelled by the light in fish-water solution produces a power loss corresponding to ~26%. In conclusion, we estimated that a combined power corresponding to less than 140 mW reaches the larva from the two sides (less than 70 mW from each side). Neither sustained calcium sensor activation nor other kinds of photodamage signs were observed during experimental imaging sessions.

### 2.5 Imaging parameters

Each imaging session started with a 5-minute control acquisition before the exposure to the convulsant. After the convulsant exposure, the imaging session proceeded alternating 5-minute acquisitions every 10 minutes for 6 cycles. In light of the phototoxicity evaluation (see Sec. "2.4 Laser power", Fig. S2 and Table S1), which was deliberately performed in harsher conditions (i.e., continuous exposure to the laser) with respect to the actual imaging configurations, this conservative choice (i.e., a discontinuous imaging protocol) was made in order to be confident not to mix or boost seizure dynamics with eventual photodamage effects arising from thermal accumulation or non-linear effects [23,31]. Therefore, for each animal we imagined 30 minutes of aberrant neuronal activity during seizure onset and propagation spanning a total time of 60 minutes.

The volumetric imaging was performed at 5 Hz with 31 stacked z-planes spanning an axial size of 150 μm. A spacing of 5 μm along the axial dimension was chosen because it coincides with the value of the half width at half maximum of the detection axial PSF. The image pixel size is ~2.2 μm × 2.2 μm and we acquired 512×512 pixel images with 16 bit depth of integer grey levels.

The sCMOS camera was operated in "synchronous readout trigger mode" to secure as long exposure time as possible. In this mode the end of each exposure coincides with the start of the next exposure and of the readout. The acquisition time (corresponding also to the exposure time in this modality) for each image is 5.8 ms, while 20 ms were allocated for the ETL flayback time.

*2.6 Zebrafish larvae*

We observed a total of 18 zebrafish (*Danio rerio*) larvae aged 4 dpf. The employed transgenic strain Tg(elavl3:H2B-GCaMP6s) [15,36], in homozygous *albino* background [37], expresses the fluorescent calcium sensor GCaMP6s under a pan-neuronal promoter with nuclear localization.

Furthermore, a subset of zebrafish larvae (n=3, Tg(elavl3:GCaMP6s), raised in 0.003% phenylthiourea (P7629, Sigma-Aldrich) to avoid pigmentation) were recorded in previous experiments [38]. Fish were maintained according to standard procedures [39] and their handling was carried out in accordance with European and Italian law on animal experimentation (D.L. 4 March 2014, no. 26), under authorization no. 407/2015-PR from the Italian Ministry of Health.

Before the imaging sessions, larvae underwent a mounting procedure as we previously described [38]. Briefly, each larva was paralyzed with a solution of d-tubocurarine (2 mM; 93750, Sigma-Aldrich) to avoid movement artefacts, included in 1.5% (w/v) low gelling temperature agarose (A9414, Sigma-Aldrich) in fish water (150 mg/L Instant Ocean, 6.9 mg/L $NaH_2PO_4$, 12.5 mg/L $Na_2HPO_4$, pH 7.2), then mounted on a custom-made glass support and immersed in fish water thermostated at 28.5 °C. After the initial 5-minute control acquisition, each larva was exposed to convulsant agent PTZ (P6500, Sigma-Aldrich) at one of the following final concentrations: 1.0, 2.5, 7.5 and 15.0 mM. We randomly assigned three larvae to each concentration. Stock solutions of PTZ were prepared by dissolving it in milliQ water, while the final concentrations used in the experiments were obtained by diluting each stock in fish water.

*2.7 Post-processing*

After the acquisitions, each time-lapse recording was manually inspected to identify movement artefacts. Where present, these artefacts were removed; in this way, from a single time-lapse movie, several motion artefact-free sub-movies were generated. The raw fluorescence images were then preprocessed as follows. First, two volumetric masks were computed as the sets of pixels whose values were smaller than two arbitrarily chosen threshold values (the chosen values are the same for all the recordings). The mask computed on the larger threshold value was then subjected to a morphology opening operation. After that, the final mask was computed as the binary-set union between the two masks. The values of all the pixels identified by the final mask were set to zero.

After the masking procedure, the 16-bit integer fluorescence images were converted in 32-bit floating point pixel-based *ΔF/F₀* signal using the following formula:

$$\Delta F / F_0 = \frac{F - F_0}{F_0 - D}, \qquad (1)$$

where $F_0$ is computed as the pixel-based first decile value along the temporal dimension, while $D$ is computed as the nearest lower integer of the lowest unmasked $F_0$ value.

Great care was taken to mount the larvae in the same position and orientation, nevertheless, in order to enhance the inter-sample consistency, we spatially aligned the brain volumes acquired from different animals in a post-processing passage. This alignment procedure was performed using a series of custom-made Python scripts that we made publicly available under the MIT License on GitHub ("https://github.com/lens-biophotonics/2P-LSFM-align"). First, time-lapse $\Delta F/F_0$ movies were spatially interpolated along the axial dimension to an almost isotropic voxel size of 2.2 μm × 2.2 μm × 2.0 μm. Then, an animal-specific rotation matrix was applied. Finally, the rotated stacks were binned to the final voxel size of 4.4 μm × 4.4 μm × 4.0 μm. In the binning procedure, solely the unmasked voxels were taken into account, so that relevant border information was not degraded. In order to generate the rotation matrices, we time-averaged one stack for each animal and then we confronted the result with a reference stack. We exploited a simulated annealing algorithm to optimize the rotation parameters, using the numerosity of the intersection set between the unmasked voxels of the two stacks as fitness value. Finally, the parameters of each matrix were fine-tuned manually by visually exploring the rotation results. For the resulting overlaid stacks, we considered as unmasked only those pixels that were unmasked in the same position in all the original stacks.

## 2.8 Voxel-based lag analysis

To perform lag analysis we computed the temporal cross-correlation (using the Pearson's coefficient) between each voxel-based $\Delta F/F_0$ trace and the global (encephalon plus spinal cord, restricted to the unmasked voxels) average $\Delta F/F_0$ trace. This procedure is justified by the fact that epileptic activity is by definition characterized by hyper-synchronicity, therefore performing the global cerebral mean makes the pathological features stand out by averaging out asynchronous activity. The voxel-based lag value is defined as the temporal cross-correlation shift associated with the maximum value of the coefficient of determination [40].

To generate single-trial voxel-based lag maps we color-mapped the lag value, limited to the unmasked voxels showing a statistically significant correlation (two-tailed test, multiplicity correction carried out with the two-stage Benjamini-Hochberg procedure, controlling the false discovery rate at 10%). We generated aggregated voxel-based lag maps by color-mapping the median values of the single-trial voxel-based lags, considering only the unmasked and statistically significant voxel of each acquisition.

Lag histograms for the relative frequencies of voxel-based lag values were computed for each acquisition and each ROI; voxel-based lag values related to the same individual event (unique occurrence of "control", "ictal" or "postictal" activity) divided in multiple sub-movies were pooled together. After that, aggregated bar plots were generated by computing the first, second and third quartiles of the distributions for each class of the histograms.

## 2.9 Statistical analysis

For region of interest (ROI)-based lag analysis, the lag value was computed for each acquisition as the median lag value among all the unmasked voxels in the ROIs that displayed also a statistically significant correlation. If the percentage of voxels in the ROI displaying a statistically significant correlation with respect to the total number of unmasked voxels was lower than 20%, then the observation was excluded. ROI-based lag values related to the same individual event (unique occurrence of "control", "ictal" or "postictal" activity) divided in multiple sub-movies were aggregated by taking their median values. ROI-based statistical analysis of lag delay was performed using Kruskal–Wallis one-way analysis of variance followed by a post-hoc pairwise multiple comparison procedure with two-tails Dunn test. Multiplicity correction was carried out with the Benjamini-Hochberg procedure controlling the false discovery rate at 5%.

Voxel-based correlation coefficients were calculated using the Pearson's coefficient as the correlation between the voxel-based $\Delta F/F_0$ temporal traces and the global (encephalon plus spinal cord, restricted to the unmasked voxels) average $\Delta F/F_0$ temporal trace. ROI-based correlation coefficients and ROI-based $\Delta F/F_0$ values were computed as the averages (restricted to the unmasked voxels of ROI) of the voxel-based respective quantities. In addition, for the analysis of the variation of the mean $\Delta F/F_0$ values over exposure time, we computed for each time point and each animal the weighted averages of the ROI-based mean $\Delta F/F_0$ values employing the number of temporal frames of the respective time-lapse as weight. For the analysis of the variation of the standard deviation of the whole-brain $\Delta F/F_0$ traces over PTZ concentration, we first computed the average $\Delta F/F_0$ value (over the whole brain) for each time point of the traces, excluding the masked voxels. Then, we computed the standard deviation of this unidimensional trace. The aforementioned weighted averaging procedure was applied also in this case. The obtained standard deviation values were finally fitted with a linear regression line with respect to the PTZ concentrations.

For the statistical analyses of the ROI-based and voxel-based correlation coefficients and of the variation of the mean $\Delta F/F_0$ values over exposure time we employed general linear mixed models [41] (GLMMs) implemented in the R [42] language in which we defined the animals as a random factor. For the analyses of the correlation coefficients relating to submaximal PTZ concentrations and of the variation of the mean $\Delta F/F_0$ values over exposure time, we set the categorical PTZ concentrations and the interaction between the categorical PTZ concentrations and the exposure time as fixed factors in the GLMMs. For the analyses of the correlation coefficients relating to maximal PTZ concentration, we set only the categorical PTZ concentrations as a fixed factor.

For the correlation coefficient voxel-based analyses, we used the frequentist package lmerTest [43] and then we corrected the results (two-tailed test) for the multiplicity using the Benjamini-Hochberg procedure, controlling the false discovery rate at 10% to test statistical significance. For the analyses of the voxel-based correlation coefficients relating to submaximal PTZ concentration, we extracted the p-values directly from the GLMMs, while for the analyses of voxel-based correlation coefficients relating to maximal PTZ concentration we employed linear contrasts based on the respective GLMM.

For the analyses of the ROI-based correlation coefficients and of the variation of the mean $\Delta F/F_0$ values over exposure time we used the Bayesian package brms [44] employing NUTS sampler (target average proposal acceptance probability during adaptation period: 0.99, maximum allowed tree depth: 15, number of total iterations per chain including warmup: 3000) and setting uninformative priors. For the analyses of the ROI-based correlation coefficients at submaximal PTZ concentration and of the variation of the mean $\Delta F/F_0$ values over exposure time, statistical significance was tested by computing the credibility intervals (CIs) at 95% for the posterior distributions of parameters. For the analysis of the ROI-based correlation coefficients at maximal PTZ concentration, statistical significance was tested employing Bayesian non-linear hypothesis testing based on the respective GLMM and setting the statistical significance threshold at 95% for the posterior probabilities.

## 3. Results

### 3.1 2P-LSFM setup optimization for fast acquisition

We specifically devised a 2P LSFM setup to perform fast whole-brain imaging in zebrafish larvae. We used double-sided illumination, thus covering the field-of-view with two partially overlapped thinner and shorter light-sheets (as performed by Truong and collaborators [24]) in place of a single, thicker, and longer one. We matched the radial dimension of the light sheets (~6 μm FWHM, Fig. S1a) with the axial detection PSF size (~10 μm FWHM), thus achieving improved axial resolution whilst illuminating the whole sample plane. Moreover, due to the two counterpropagating infrared beams, this excitation geometry provides homogeneous illumination and strong attenuation of striping artefacts [45,46]. Noteworthy, the typical size of the larval brain along the lateral dimension (~400 μm) results slightly smaller than the beam effective confocal parameter (Fig. S1b). In particular, the radial dimension of the light-sheets remains lower than 7 μm over a path longer than 200 μm, i.e. corresponding to the brain-half related to the respective illumination emi-field, while also maintaining uniform illumination over this distance (Fig. S1c).

In 2P excitation, fluorescence emission depends on the square of light intensity, hence it is important to maximize the excitation peak power. Therefore, we employed an EOM and a polarizing beam splitter (that steers the light depending on the instantaneous light-polarization state induced by the EOM operation) to alternatively convey excitation light in either of the two illumination arms, thus maximizing the peak power of excitation. It should be noted that the employed EOM operating frequency is three orders of magnitude larger than the camera acquisition frequency, therefore no blinking can be observed in the recordings. This arrangement can ideally produce a 2-fold increase of the fluorescent signal compared to halving the peak power to simultaneously illuminate both sides, e.g. using a non-polarizing beam splitter. In fact, in the first case the peak power is preserved, albeit the illumination time is halved, the latter having a linear effect on the decreasing of the signal levels. On the contrary, in the second case the peak power is halved, inducing a quadric decrement in the signal that would not be compensated by the preservation of the total illumination time.

In order to experimentally test this effect, we measured the intensity of the fluorescence light produced in a fluorescein solution using a polarizing or a non-polarizing beam-splitter (Fig. S1c,d) while keeping a constant excitation power. We indeed observed an increase of 84% in the generated signal when using the polarizing beam-splitter, confirming the validity of our approach. A large part of the discrepancy with respect to the theoretical case (100% increase) is to be ascribed to residual polarization distortions. Indeed, when the polarization is switched, about 2.5% of residual illumination intensity can still be measured in the original optical path and this accounts for 10 percentage points of the difference.

Furthermore, the setup employs a quarter-wave and two half-wave retarders to make the polarization plane of the excitation light parallel to the light-sheet plane. By photoselecting a subpopulation of fluorophores emitting preferentially towards the detection objective, this implementation drastically increases the signal levels in 2P LSFM, as we showed in a previous article [47].

We maximized fluorescent signal collection using a high-NA detection objective. Its focus was remotely displaced to match it with the position of the illumination light-sheet, using an ETL as described by Fahrbach *et al.* [48]. This allowed fast volumetric imaging while keeping the bulky objective still, which would have otherwise produced detrimental pressure waves impinging on the sample.

The positioning of the ETL immediately before the camera-side objective pupil (a configuration similar to the one adopted in Ref. [48]) introduces a small degradation of optical telecentricity, i.e. an axial-position dependent magnification component. The magnification alteration at the two extremes of the range commonly used in our acquisitions (150 μm) was quantified and it was found to be below 7% with respect to the magnification at the center of the range (a level similar to what described by Fahrbach *et al.* [48]). Therefore, considering the functional (i.e. not structural) scope of our observations, this effect is deemed negligible.

The image generated by the detection objective tube lens is demagnified to a final 3× (i.e. radically different from the original magnification of the sample-side objective: 20×) onto the camera sensor by an optical relay system (composed by a second couple of objective and tube lens), thus optically binning to concentrate light. This design allows for optimal fluorescence collection using a high-NA detection objective without compromising on wide field-of-view. This enlargement of the field-of-view comes at the expense of optical aberrations in the field periphery. However, these aberrations are still acceptable in the area occupied by the sample, especially considering the functional nature of our observations. In order to achieve these results, care was taken to match the NA of the sample-side objective ($NA_S$) with that ($NA_C$) of the camera-side objective: $NA_S M \approx NA_C$, where $M$ is the lateral magnification. If this condition is verified, then the optical system can exploit the full light-collection capabilities of the high-NA sample-side objective. Noteworthy, in this configuration the ETL (placed near the back aperture of camera-side objective) does not limit the NA of the detection system, since its aperture (16 mm) is larger than the magnified image of the objective pupil (12 mm).

Furthermore, the demagnification adopted allows the image to be focused onto a small central portion (512 × 512 pixels) of the sCMOS camera sensor, providing enhanced number of photons collected by each camera pixel and increased image readout frequency. We chose an image pixel size of 2.2 μm in order to sample neuronal nuclei (~5 μm) according to the Nyquist principle.

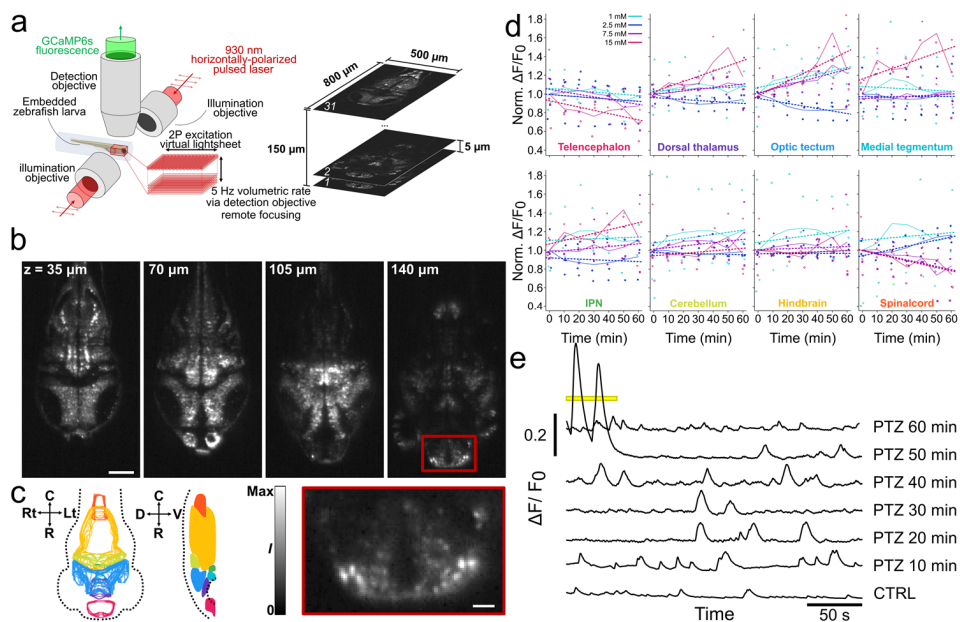

Fig. 2. Fast volumetric functional imaging of seizures in zebrafish larvae. (a) Left: close-up schematic showing the axial scanning direction, the excitation light polarization plane (double-headed arrows), and the geometric disposition of the larva and the objectives. Right: scheme showing the parameters of the volumetric acquisition process. (b) Top: four coronal sections showing temporal average intensity projections at different dorso-ventral depths of a larva (indicated on the panels, with respect to the dorsal surface). Scale bar: 100 μm. Orientation as in (c), left panel. Bottom: magnification of the area indicated by the red rectangle. Scale bar: 20 μm. (c) Outlines of the different ROIs (in different colors) considered in the analysis; left: coronal projection, right: sagittal section over larval midline. R: rostral, C: caudal, D: dorsal, V: ventral, Rt: right, Lt: left. ROI names reported in (d). (d) Scatter plots of the mean values of the $\Delta F/F_0$ traces as a function of the PTZ exposure time. Different colors and shapes indicate different PTZ concentrations (as specified in the legend) and animals, respectively. A small amount of jitter was applied on the x-axis to improve visualization. The traces were generated by averaging the activity signal over different ROIs, as indicated on the plots (in the same colors as in (c)). Predicted values generated by the Bayesian model for each PTZ concentration are indicated as dashed lines with corresponding colors. Time-point-based averages for each concentration are linked by solid lines of the corresponding colors (all values were normalized on the mean value of the corresponding control acquisitions at time zero). IPN: interpeduncular nucleus. Number of larvae measured: 3 for each condition tested. (e) Examples of $\Delta F/F_0$ traces from a single larva exposed to 15 mM PTZ, recorded during different acquisition times as indicated on the right. Traces were generated by averaging the activity signal over the whole brain. The yellow bar indicates an individual ictal event that happened 50 minutes after PTZ exposure.

### 3.2 Brain-wide PTZ-induced seizures mapping

The system allowed us to capture neuronal activity in the intact whole brain of zebrafish larvae expressing the calcium indicator GCaMP6s [15,36] once every 200 ms (Fig. 2a), yet with cellular resolution in the vast majority of the cerebral tissue (Figs. 2b, S3 and S4). Despite the counterpropagating-beams architecture can be prone to a partial laser blocking by the larval eyes, with respect to more complex illumination strategies [15,28], the reduced image quality in the deepest portion of the forebrain did not prevent us from extracting meaningful functional activity even from this area (Fig. S5). Notably, we exploited the fast volumetric imaging abilities of the setup and its invisible nonlinear excitation to investigate on a brain-wide scale the neuronal dynamics occurring during seizures in a photosensitive system such as an epileptic brain. We induced seizures of different entities in 4 dpf transgenic zebrafish larvae using different concentrations (1.0, 2.5, 7.5 and 15.0 mM) of PTZ, a $GABA_A$ receptor antagonist widely used as a convulsant drug [38,49]. Analogously to what we already described [38], we observed a linear relationship between the PTZ concentration and the global brain calcium activity (Fig. S6). The thorough brain mapping provided by the system allowed for fine monitoring of seizure onset and propagation across multiple anatomic districts spanning the entire brain (Fig. 2c), identified according to available larval zebrafish brain atlases [50,51].

In order to quantify the effects of the different PTZ concentrations tested on neuronal activity amplitude in different brain regions over time, we draw the scatter plots shown in Fig. 2d, showing the results obtained from 3 larvae for each of the four PTZ concentrations.
We found that larval brain regions are differently recruited during seizures. Indeed, at 1.0 mM PTZ we did not observe statistically significant alterations. The same at 2.5 mM PTZ, with the exception of a statistically significant (the 95%-CI excludes the zero value) increase on the relative (with respect to the control condition) mean $\Delta F/F_0$ value in the spinal cord (95%-CI: [0.42; 6.24]$\times 10^{-3}$ min$^{-1}$).

This scenario dramatically changes at the higher concentrations tested. At 7.5 mM PTZ concentration we observed statistically significant decreases for the telencephalon (95%-CI: [-4.82; -0.50]$\times 10^{-3}$ min$^{-1}$) and the spinal cord (95%-CI: [-6.47; -0.52]$\times 10^{-3}$ min$^{-1}$). Finally, when exposed to saturating PTZ concentration (15 mM), zebrafish larvae undergo an extensive functional connectivity rearrangement marked by an overall increase in neuronal activity and synchronicity (Sect. 3.3). In this case, we observed statistically significant decreases for the telencephalon (95%-CI: [-5.59; -1.19]$\times 10^{-3}$ min$^{-1}$) and the spinal cord (95%-CI: [-6.49; -0.72]$\times 10^{-3}$ min$^{-1}$) and statistically significant increases for the optic tectum (95%-CI: [3.35; 12.48]$\times 10^{-3}$ min$^{-1}$), dorsal thalamus (95%-CI: [2.12; 9.05]$\times 10^{-3}$ min$^{-1}$), medial tegmentum (95%-CI: [1.47; 9.89]$\times 10^{-3}$ min$^{-1}$) and interpeduncular nucleus (95%-CI: [0.60; 7.06]$\times 10^{-3}$ min$^{-1}$).

This functional rearrangement eventually leads to the appearance of acute seizure ictal-like events [52,53], depicted in Fig. 2e, which shows typical whole-brain $\Delta F/F_0$ traces, observed during pre-exposure recording (CTRL) and after exposure to maximal concentration of PTZ. After an initial increase in amplitude and frequency of global brain activity, between 10 and 40 minutes upon the addition of the convulsant drug, the scenario abruptly changes with the appearance of typical high-amplitude and low frequency ictal events (yellow highlight at 50 min).

### 3.3 Hyper-synchronicity during seizure activity

In order to characterize how the degree of hyper-synchronicity in neuronal activity varies during the development of seizure, we studied the trend of the correlation coefficient between the voxel-based calcium traces and the global average calcium trace. We observed that some of the brain regions exhibited similar neuronal behaviors, therefore here we mainly focused our analysis on four regions distributed over the whole extension of the encephalon, summarizing the globality of the observed neuronal behaviors: telencephalon, optic tectum, dorsal thalamus and spinal cord. Aggregate results from all the larvae are shown in Fig. 3. We analyzed the data from animals exposed to maximal and submaximal concentrations of PTZ separately, since we have shown that only the former are characterized by the emergence of ictal activity, i.e. paroxysmal activity typical of acute seizures.

For submaximal concentrations, we characterized the temporal dependence of the correlation coefficient values on the exposure time and we generated color maps based on this effect [17], quantified as t-value (i.e. the computed value of the Student's t-statistic), while checking its statistical significance. As shown in Fig. 3a, almost no statistically significant effects are observed at 1.0 mM PTZ concentration, while at 2.5 mM PTZ concentration we observed a significant negative correlation for few voxels, mostly located in the spinal cord in addition to a small number of others dispersed (but not isolated) throughout the encephalon. Conversely, at 7.5 mM PTZ concentration we observed the emergence of a widespread positive correlation between the voxel-based calcium traces and the global average trace, meaning that the value of the correlation coefficient increases with the exposure time. The increased correlation among neuronal activity is the hallmark of seizures, even when ictal activity is not yet present. It is interesting to note that some subregions show an unexpected appearance of negative correlation, such as portions of the spinal cord and part of the medial tegmentum, meaning that for these subregions the value of the correlation coefficient actually decreases with the increasing of the exposure time.

We then performed the same correlation analysis at ROI level and no statistically significant effect was visible for 1.0 mM and 2.5 mM PTZ concentrations (Fig. 3b). On the contrary, for the 7.5 mM PTZ concentration, the correlation coefficient increased during exposure time for ROIs such as the dorsal thalamus (95%-CI: $[2.47; 9.56] \times 10^{-3}$ min$^{-1}$), the optic tectum (95%-CI: $[4.29; 10.97] \times 10^{-3}$ min$^{-1}$), the cerebellum (95%-CI: $[3.99; 9.96] \times 10^{-3}$ min$^{-1}$) and the hindbrain (95%-CI: $[1.37; 7.74] \times 10^{-3}$ min$^{-1}$) in a statistically significant manner (the 95%-CI excludes the zero value). The decrease observed for the spinal cord ROI, however, did not result as significant (95%-CI: $[-4.67; 3.34] \times 10^{-3}$ min$^{-1}$); this lack of significance can be explained by the fact that the decrease in the correlation coefficient detected for small subregions included in this ROI is washed out when taking into account the whole region.

In larvae exposed to the maximal PTZ concentration (15 mM) we computed the correlation value differences among ictal, postictal and control conditions and the aggregate results for the voxel-based analyses are shown in Fig. 3c. We observed a significant and widespread increase in the correlation coefficient in the ictal condition when compared to control or postictal conditions in the whole encephalon. On the contrary, we observed a reduced number of voxels showing significant differences between the postictal and control conditions in the encephalon, the majority of them exhibiting an increase in the correlation coefficient and being concentrated in subregions belonging to the inner parts of midbrain and hindbrain. Interestingly, the spinal cord behaved differently, showing almost no differences between ictal and control conditions, while a significant decrease in the correlation coefficient can be observed limited to its most dorsal subregion in the postictal condition compared to the control condition. Finally, in the comparison between the ictal and the postictal conditions, the spinal cord behaved similarly to the encephalon, showing a widespread increase in the correlation coefficient.

We characterized these differences at ROI level and the results are shown in Fig. 3d. We observed an increase in the correlation coefficient for the ictal condition compared to the control or the postictal conditions for all the ROIs belonging to the encephalon, e.g. for the telencephalon (+0.28 increment, 90%-CI: +0.18, +0.38, posterior probability ictal > control: ~ 100% and +0.34 increment, 90%-CI: +0.26, +0.42, posterior probability ictal > postictal: ~ 100%, respectively), the optic tectum (+0.53 increment, 90%-CI: +0.36, +0.70, posterior probability ictal > control: ~ 100% and +0.38 increment, 90%-CI: +0.25, +0.51, posterior probability ictal > postictal: ~ 100%, respectively) and the dorsal thalamus (+0.55 increment, 90%-CI: +0.42, +0.69, posterior probability ictal > control: ~ 100% and +0.41 increment, 90%-CI: +0.31, +0.51, posterior probability ictal > postictal: ~ 100%, respectively) ROIs. Conversely, in the comparison between the postictal and the control conditions we observed only a single significant difference in the correlation coefficient among the ROIs belonging to the encephalon: an increase for the dorsal thalamus ROI (+0.14 increment, 90%-CI: +0.03, +0.26, posterior probability postictal > control: 98%). Remarkably, we did not observe significant differences for the spinal cord ROI comparing the ictal or the postictal conditions to the control condition (+0.13 increment, 90%-CI: -0.04, +0.29, posterior probability ictal >

control: 89% and -0.02 decrement, 90%-CI: -0.17, +0.12, posterior probability postictal < control: 59%, respectively), while we observed a significant increase (+0.15 increment, 90%-CI: +0.02, +0.29, posterior probability ictal > postictal: 97%) in the correlation coefficient for the ictal condition compared with the postictal condition.

In summary, the exposure to maximal PTZ concentration (15 mM) induced a general increase in neuronal activity and synchronicity, except for the most caudal and rostral regions (namely spinal cord and, limited to the neuronal activity, telencephalon).

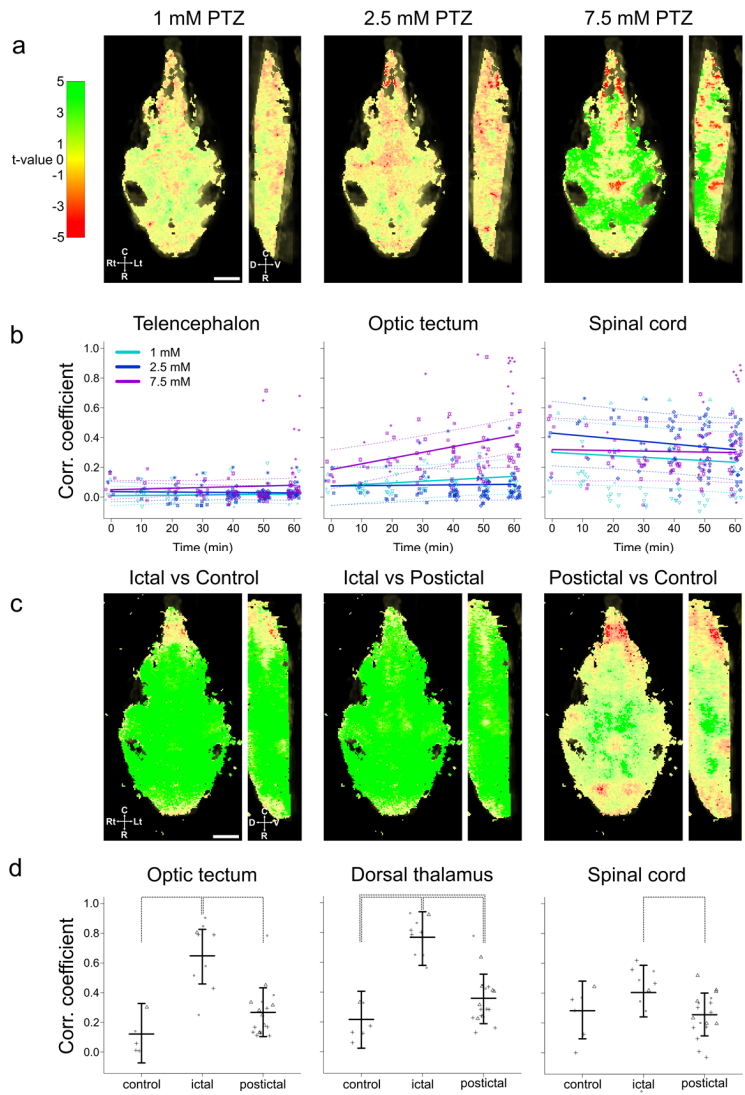

Fig. 3. Correlation analysis. (a) Color maps (coronal and sagittal sections) depicting the effects of PTZ concentration and exposure time on the correlation coefficient value computed between single-voxel calcium activity and global calcium activity. These maps represent the aggregate results from all the larvae exposed to sub-maximal PTZ concentrations. Masked voxels are depicted in grey values, non-statistically significant voxels in unsaturated colors and statistically significant voxels in saturated colors. The voxel-based t-value associated to the effect of the interaction between PTZ concentration and exposure time is color-mapped as described by the color-bar on the left. R: rostral, C: caudal, D: dorsal, V: ventral, Rt: right, Lt: left. Scale bar: 100 μm. (b) Scatter plots showing the distribution of the ROI-based (as indicated on the top) correlation coefficients (on the y-axis) as a function of PTZ concentration (different colors) and exposure time (on the x-axis). Data related to different animals are indicated by different shapes. A small amount of jitter was applied on the x-axis to increase readability. Solid lines indicate predicted values generated by the Bayesian model, while dotted lines indicate the 95%-CIs. The regression coefficient related to the optic tectum and the 7.5 mM PTZ exposure resulted as statistically significant. (c) Color maps (coronal and sagittal sections) depicting the effects of the control, ictal and postictal conditions on the correlation coefficient value. These maps represent the aggregate results from all the larvae exposed to maximal PTZ concentration. The colors map voxel-based t-values associated to the effect of the condition. Color code, color saturation, grey values and scale bar as in (a). (d) Scatter plots showing the distribution of the ROI-based (as indicated on the top) correlation coefficients (on the y-axis) at maximal PTZ concentration in control, ictal and postictal conditions (on the x-axis). Data related to different animals are indicated by different shapes. A small amount of jitter was applied to the x-axis to increase readability. The horizontal lines indicate predicted values generated by the Bayesian model and their error bars indicate the 95%-CIs. Statistically significant differences are indicated by dotted lines.

*3.4 CRIWs: a postero-anterior seizures propagation pattern*

During this phase of paroxysmal activity, appearing only in larvae exposed to the convulsant maximal concentration, we observed the emergence of previously unreported fast rhythmic ictal waves propagating in postero-anterior direction, which we termed caudo-rostral ictal waves (CRIWs). Visualization 1 (Scale bar: 100 μm) reports a volumetric recording of an individual CRIW event shown as a selected subset of coronal sections. To characterize the spatio-temporal propagation features of CRIWs, we calculated the temporal delay of each voxel activation with respect to the global activity (Fig. 4a and Visualization 2). In order to do so, we computed the cross-correlation between the voxel-based and the global average activities. In this way, by identifying the voxel-based maxima in the temporal cross-correlation function (i.e. the "lag value"), it is possible to visualize the spatial spreading path of the seizure activity.

While the individual CRIW initiation appears to be synchronous all over the larval brain, its propagation dynamics assume a wave pattern consistently starting in the most caudal regions (spinal cord, hindbrain) and spreading then to the most rostral ones (optic tectum, telencephalon) as shown in Fig. 4b. This seizure propagation pattern is characterized by high-amplitude and long-lasting calcium transients (Fig. 4b) that are typical of ictal activity since it is not observed during physiological activity (Fig. 4c). Importantly, this wave pattern does not emerge merely as an aggregate average. We verified the across-event consistency of the aggregated data of the ictal phase by assessing the concordance of the voxel-based lag-values with those of all the recorded ictal events from which the aggregated map is derived (Fig. S7). Moreover, this recurrent propagation motif is also visible in single-trial data (Fig. S8a). In the post-ictal phase—a state characterized by a marked overall depression of neuronal activity (Fig. 2e) and a decreasing synchronicity trend (Fig. 3c,d)—we report a wave-like moving pattern too, albeit less pronounced (Fig. 4c).

We then quantified lag value differences among brain regions (Fig. 4d). Interestingly, despite the entire CRIW typically lasting a few seconds, it consistently spans the entire brain, from the most caudal regions to the most rostral ones, in about 1 second, a time-frame inaccessible to previous 2P light-sheet setups with lower acquisition frequency [16,17]. Finally, computing the distribution of voxel-based lag values for each brain district (Figs. 4e and S8b), we confirmed that different brain regions are differently recruited during the occurrence of seizures.

The statistical analysis confirmed the existence of significant (p-value = 0.0002) differences among several rostral and caudal regions in the ictal condition, as shown at the bottom of the plot. In particular, the q-values for post-hoc Dunn test were observed to be less than 0.05 for the following comparisons: for telencephalon versus hindbrain (q-value = 0.0122) and spinal cord (q-value = 0.0085) ROIs, for optic tectum versus interpeduncular nucleus (q-value = 0.0358), hindbrain (q-value = 0.0084) and spinal cord (q-value = 0.0088) ROIs and for dorsal thalamus versus hindbrain (q-value = 0.0159) and spinal cord (q-value = 0.0099) ROIs. Few of these differences were also observed in the postictal condition (p-value = 0.0032). In particular, the q-values for post-hoc Dunn test were observed to be less than 0.05 for the following comparisons: for optic tectum versus medial tegmentum (q-value = 0.0272), hindbrain (q-value = 0.0241) and spinal cord (q-value = 0.0009) ROIs. Finally, no significant differences were observed in the control condition (p-value > 0.05).

A recent work [54], employing zebrafish and PTZ as a paradigm (5-6 dpf, 10 mM PTZ) to study convulsion-related activity with spinning disk confocal imaging, showed a specific seizure propagation pattern moving in antero-posterior direction (with ictal events starting in the telencephalon and propagating caudally, up to the cerebellum). Since the propagation pattern described in [54] is opposite to the one we observed in CRIWs, we decided to perform the following investigations as a control.

First, to validate the results of our analysis, we re-processed a subset of our data pertaining to ictal activity with the method described in [54]. With this procedure lag values are computed on the basis of the voxel-based local maxima of the time derivative of the smoothed (over 2 s) $\Delta F/F_0$ trace. Representative lag maps resulting from this method are shown in Fig. S9 and they clearly highlight a (albeit noisier) caudo-rostral propagation pattern of activity (i.e. the CRIWs). This result validates the appropriateness of our analysis method to identify and describe the lag of activation in our measurements.

Then, in order to exclude the possibility of CRIWs to be determined by a combination of PTZ saturating condition and some 2P imaging effect (e.g. heating), we applied our validated lag analysis method to 5 Hz wide-field fluorescence measurements of zebrafish larvae neuronal activity, performed in our own experimental conditions (4 dpf, 15 mM PTZ). Indeed, 1P wide-field fluorescence imaging, employing a collimated LED source to illuminate the sample, is a technique much less prone to photobleaching, phototoxicity and photodamage with respect to other fluorescence microscopy techniques. Figure S10 shows the lag map during ictal activity and it highlights a seizure propagation pattern consistent with CRIWs (particularly taking into account the lack of depth discrimination in wide-field fluorescence measurements), meaning that the phenomenon we described does not depend on the optical method used to observe it.

Finally, we performed 2P LSFM imaging of larval brain activity in the experimental conditions described in Ref. [54], namely 6 dpf larvae and 10 mM PTZ. Figure S11 shows the median lag map of the ictal activity highlighting a propagation pattern different both from CRIWs and from the one previously described in Ref. [54]. Similarly to Liu and Baraban, we revealed an early activation of the telencephalon. However, in contrast with what they described, this activation is concurrent with the one of the spinal cord and of part of the optic tectum. The remaining part of the optic tectum shows a late activation along with the habenulae and the superficial part of the hindbrain.

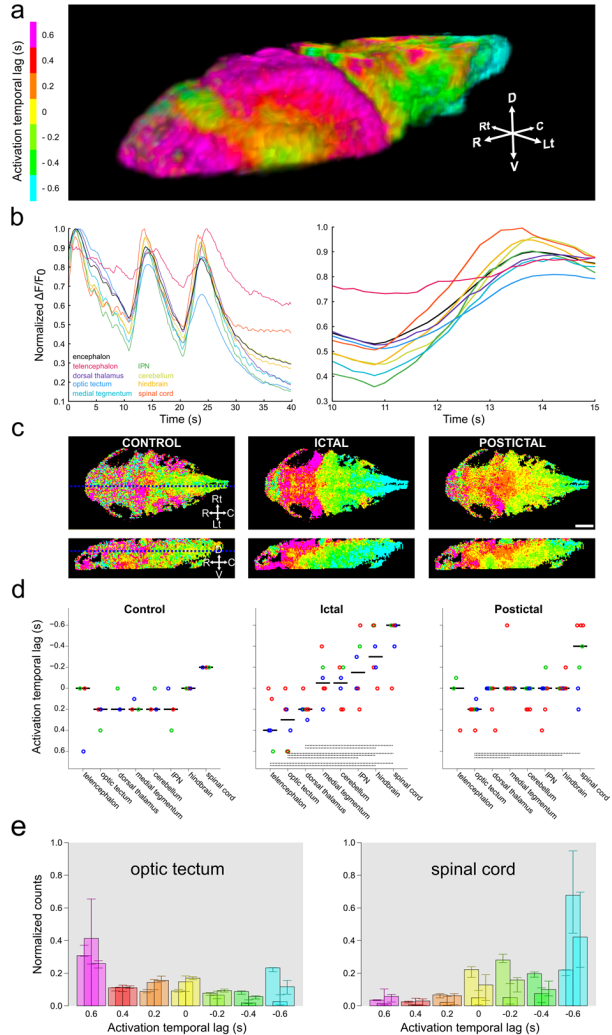

Fig. 4. Caudo-rostral ictal waves (CRIWs) during severe seizures in zebrafish larvae. (a) 3D rendering of the lag map (reported in (c)) showing the voxel-based temporal position of the maximum of the cross-correlation function (coefficient of determination) between the voxel-based calcium signal and the global cerebral average during the ictal phase. This lag map represents the aggregate results from larvae exposed to PTZ maximal concentration (n. larvae = 3, n. CRIWs = 6). The lag value is color-coded as specified by the color bar. R: rostral, C: caudal, D: dorsal, V: ventral, Rt: right, Lt: left. See Visualization 2 for a rotating rendering of this lag map (Scale bar: 100 μm). (b) $\Delta F/F_0$ signal from a single trial acquired during the ictal phase for different ROIs (indicated by different colors as specified by the legend and corresponding to Fig. 2c). The signals were normalized dividing them by the respective maximum values. The lag map corresponding to these traces is shown in Fig. S8a. Magnification in the inset. (c) Lag maps for control (n=3), ictal (n=6) and postictal (n=7) events observed in the 3 larvae exposed to PTZ maximal concentration. Top row: coronal sections; bottom row: sagittal sections. The sectioning planes are indicated as dashed blue lines on the first panels. Color code and abbreviations as in a. Scale bar: 100 μm. (d) Scatter plots of the median lag values for each condition (as indicated in the plots) and ROI (on the x-axis). Different colors indicate different larvae. Each point corresponds to a single event. A small amount of jitter was applied on the x-axis to improve readability. Horizontal solid lines indicate the median value for each ROI. Statistically significant differences are indicated by horizontal dotted lines at the bottom of the graphs. (e) Bar plots showing the relative frequencies of the voxel-based lag values for two ROIs (as indicated in the plots). The bar colors correspond to the color code used in (a) and (b). The height of each bar corresponds to the median value of the single-acquisition-based frequency distribution, while the error bars indicate the first and the third quartiles of the distribution. The first, second and third bars in each lag class correspond to the control, ictal and postictal situations, respectively. The plots for remaining ROIs are shown in Fig. S8b.

## 4. Discussion

Here, we describe a novel 2P LSFM setup employing pulsed infrared excitation light to perform high-speed cellular-resolution imaging of the larval zebrafish brain. Compared to conventional one-photon LSFM (employing excitation in the visible range that can induce unwanted visual stimulations [16]), 2P LSFM is prone to low SNR issues—owing to low 2P absorption probability—which typically is the limiting factor toward rapid volumetric imaging. Hence, in order to push further the volumetric temporal resolution of the system, we adopted a series of technical implementations to optimize the setup for maximum SNR. Particularly, using an EOM we alternated the illumination of the two halves of the field-of-view to maximize the peak power of excitation light. This, due to the nonlinear nature of 2P absorption, led to a more than 80% increase in signal with respect to halving the peak excitation power between the two illumination arms. Moreover, as recently demonstrated by our group [47], controlling the excitation light polarization to orient the emission toward the detection objective led to a large increase in fluorescence collection. We employed a remote focusing approach, using an ETL in the detection path, to enable volumetric signal collection by a high-NA immersion lens, while keeping the objective still (thus avoiding the generation of disturbing pressure waves on the sample by the high frequency oscillation). Finally, we performed an optical binning of the final image on the detector sensor to decrease readout and integration times while concentrating the fluorescence signal. Collectively, the adopted implementations allowed for system SNR boosting, enabling to quintuplicate (5 Hz, 600 × 800 × 150 μm$^3$, 31 sections) the volumetric acquisition rate of state-of-the-art brain-wide LSFM imaging using non-linear excitation. Indeed, previous applications of 2P LSFM to larval zebrafish functional imaging yielded to 1 volume/s mapping at limited brain depths (9-10 sections over 64 to 100 μm) [16,17,27] or to wider mapping (250 μm depth, 52 sections) at the cost of volumetric sampling rate (0.5 Hz) [28]. Actually, in the presented setup the main limitation factor for acquisition frequency is represented by the dynamics of the large-aperture ETL. Indeed, the single-image acquisition time (5.8 ms, i.e. the integration time of each plane in the volumetric scansion) is very close to the ETL response time (5 ms). Further improvements in temporal resolution may stem from fine-tuning the analog waveform driving the ETL, in order to optimize its dynamics behavior, as shown in [48].

In recent years, a few works have performed one-photon LSFM functional imaging of the larval brain to explore network dynamics during seizures, despite not thoroughly exploiting the fast real-time volumetric access provided by LSFM architecture [55–57]. We employed our fast 2P LSFM system to investigate on a brain-wide scale, yet with cellular resolution, the neuronal dynamics occurring in the larval zebrafish central nervous system during acute seizures. We exposed larvae to PTZ, a widely employed convulsant that leads to uncontrolled paroxysmal activity typical of epileptic seizures by progressively blocking the inhibitory tone. We characterized brain-wide neuronal activity both in terms of amplitude and synchronicity: the hallmarks of epileptic discharges.

Exposing larvae to different PTZ concentrations, we found brain regions to be differentially recruited during seizures. Indeed, at the lowest dose (1 mM) no significant effects were observed on the activities of all brain regions considered, neither on their amplitudes nor on their synchronicity levels with respect to control, confirming what we observed in a previous work [38]. At 2.5 mM PTZ, the spinal cord shows a significant increase in amplitude over exposure time which is accompanied by a decrease in correlation with the global brain activity. Interestingly, the spinal cord shows an unexpected concentration-dependent dichotomic effect of PTZ on its functional connectivity. Indeed, when exposed to the highest concentrations (7.5 and 15 mM) we observe an opposite behavior in terms of amplitude.

At the higher concentrations (7.5 and 15 mM PTZ), we observe that midbrain districts (dorsal thalamus, optic tectum, medial tegmentum and IPN) are more susceptible to the effects of the convulsive agent. At 7.5 mM PTZ, a great part of larval encephalon positively contributes to global brain activity, whereas several voxels in the medial tegmentum and in the hindbrain show reducing correlation. At saturating PTZ concentrations (15 mM), the larval brain undergoes a profound functional connectivity rearrangement eventually leading to the appearance of ictal-like neuronal activity, as previously described [32,33,38]. Ictal activity is characterized by high-amplitude and long-duration calcium transients, confirming what previously observed using different techniques [38,53,54,58]. In this condition, we observe an even more extensive increase in correlation, involving the entire brain during ictal events, as previously reported [38,53,54,59]. Indeed, at maximal PTZ concentration all brain regions start acting as a synchronous system, eventually leading to severe paroxysmal activity represented by ictal events. Each ictal event is followed by postictal activity, a state characterized by a marked overall depression of neuronal activity. Particularly, in the postictal phase we observed a decreased synchronicity in a subregion of the spinal cord, likely underlying the ataxic phase observed during postictal behavior [32].

Through lag analysis (which we validated by comparing the results with a previously described method) we reveal that ictal activity spreads in a peculiar and consistent wave pattern spanning the entire larval brain from the most caudal regions to the more rostral ones and for this reason we named this phenomenon caudo-rostral ictal wave. The consistency of CRIWs as a pathological propagation pattern, and its independence from the optical method used to investigate it, are confirmed by its observation in measurements performed in the same experimental conditions (4 dpf larvae, 15 mM PTZ) yet employing a 1P wide-field fluorescence microscope.

This previously unreported seizure propagation pattern is characterized by high-amplitude and long-lasting calcium transients which are realistically generated by thousands of closely paced action potentials. Indeed, in this scenario, where all the inhibitory GABAergic synapses are dimmed by high PTZ concentrations, excitatory drive is predominant. Firing of neurons, normally curbed by the inhibitory tone, may trigger other neuronal populations to fire, generating an avalanche effect which determines the seizure propagation we observe.

A recent work by Liu and Baraban [54], employing 1P confocal spinning disk microscopy to image on a single brain plane 5-6 dpf larvae exposed to 10 mM PTZ, showed a seizure propagation pattern opposite to CRIWs (i.e., in an antero-posterior direction). In order to assess whether this discrepancy might be due to the different larval ages (4 *versus* 5-6 dpf) and/or to the different convulsant concentration (15 *versus* 10 mM) we used our 2P LSFM setup to map seizure dynamics in the experimental conditions reported in Ref. [54]. Differently with what we observed using 15 mM PTZ, we noticed that not all the larvae we imaged underwent severe seizures resulting in ictal events, meaning that 10 mM is a not saturating concentration of the convulsant. From lag analysis performed on ictal activity we indeed revealed an early activation of telencephalon, in accordance with what described by Liu and Baraban. However, the resulting propagation does not resemble a wave pattern, given the concurrent early activation of not contiguous brain regions. We think that this discrepancy between the two results (namely the lack of a clear wave pattern in our replicated experiment) may be explained with the particularities inherent to the diverse imaging techniques employed. The distinct propagation pattern we observed (namely CRIWs—at 4 dpf and 15 mM PTZ—and early rostral activation in our replication of Liu and Baraban conditions—at 6 dpf and 10 mM PTZ—) suggests that larval age (and the consequent brain structural and functional grade of development) plays a crucial role in the propagation of pathological neuronal activity in zebrafish. Even if we expect the propagation pattern to be less affected by the concentration of PTZ needed to effectively trigger ictal activity, we still cannot exclude an additional influence of this parameter.

The presented technique, which we successfully demonstrated in a delicate model such as an unstable epileptic brain, might be applied in the future also to other experimental models in which unwanted visual stimulation could be detrimental to unveil the neural dynamics underlying specific brain states [60] and behaviors, such as circadian rhythm [30] or virtual-reality driven responses. Finally, the temporal resolution achieved by the system, combined with its cellular resolution, will enable a future extension of the setup with an optogenetic excitation control module, to perform real-time closed-loop control of neuronal activity. For example, compared to post-processing interpolation methods—which could be used to upscale the temporal resolution in the calculation of lag values based on non-trivial inferences on *in-vivo* non-linear dynamics of the calcium sensor—our native 5 Hz whole-brain sampling frequency will allow future experiments involving real-time optogenetic suppression of ongoing seizure activity based on closed-loop feedback approaches.

## 5. Conclusions

In conclusion, we devised a microscope that enables high-speed volumetric functional imaging of the entire larval zebrafish brain, while employing nonlinear excitation. The combination of multiple technical implementations allowed us to boost fluorescence signal collection in 2P LSFM, thus quintuplicating the temporal resolution of state-of-the-art 2P LSFM, while keeping low the laser intensity on the specimen. The setup allowed the onset and propagation of seizures to be studied in a non-invasive fashion in a highly sensitive system such as an epileptic brain. Owing to the volumetric rate reached (5 Hz), we could observe and spatio-temporally describe for the first time the emergence of CRIWs: fast travelling rhythmic ictal waves propagating in caudo-rostral direction. The system we devised enabled us to deepen the knowledge of whole-brain dynamics and functional connectivity reorganization during acute seizures in zebrafish larvae.

**Acknowledgements.** This project has received funding from the European Research Council (ERC) under the European Union's Horizon 2020 research and innovation programme (grant agreement No 692943).

We thank Dr. Vladislav Gavryusev, Dr. Luca Pesce (both from University of Florence and the European Laboratory for Non-Linear Spectroscopy, LENS) and Dr. Chiara Fornetto (from LENS), for their help with closed-loop galvanometric mirror PID controller configuration, detection-PSF measurements and phototoxicity measures, respectively. We thank Riccardo Ballerini and Mauro Giuntini, respectively from mechanical and electronic workshops at LENS, for technical support and custom pieces production. Moreover, we thank Dr. Anna Letizia Allegra Mascaro (from LENS and National Research Council), Dr. Francesco Resta and Dr. Alessandro Scaglione (both from LENS and University of Florence) for their useful insights and suggestions. We acknowledge Lizzy Griffiths for zebrafish larva drawing.

**Disclosures.** The authors declare no conflicts of interests.

**Data Availability.** Data underlying the results presented in this paper are not publicly available at this time but may be obtained from the authors upon reasonable request.

See Supplement 1 for supporting content.

**Bibliography**

1. F. Helmchen and W. Denk, "Deep tissue two-photon microscopy," Nature Methods **2**, 932–940 (2005).
2. D. A. Dombeck, A. N. Khabbaz, F. Collman, T. L. Adelman, and D. W. Tank, "Imaging Large-Scale Neural Activity with Cellular Resolution in Awake, Mobile Mice," Neuron **56**, 43–57 (2007).
3. P. J. Keller, A. D. Schmidt, J. Wittbrodt, and E. H. K. Stelzer, "Reconstruction of Zebrafish Early Embryonic Development by Scanned Light Sheet Microscopy," Science **322**, 1065–1069 (2008).
4. N. J. Sofroniew, D. Flickinger, J. King, and K. Svoboda, "A large field of view two-photon mesoscope with subcellular resolution for in vivo imaging," eLife **5**, e14472 (2016).
5. D. G. Ouzounov, T. Wang, M. Wang, D. D. Feng, N. G. Horton, J. C. Cruz-Hernández, Y.-T. Cheng, J. Reimer, A. S. Tolias, N. Nishimura, and C. Xu, "In vivo three-photon imaging of activity of GCaMP6-labeled neurons deep in intact mouse brain," Nature Methods **14**, 388–390 (2017).
6. T.-W. Chen, T. J. Wardill, Y. Sun, S. R. Pulver, S. L. Renninger, A. Baohan, E. R. Schreiter, R. A. Kerr, M. B. Orger, V. Jayaraman, L. L. Looger, K. Svoboda, and D. S. Kim, "Ultrasensitive fluorescent proteins for imaging neuronal activity," Nature **499**, 295–300 (2013).
7. H. Dana, Y. Sun, B. Mohar, B. K. Hulse, A. M. Kerlin, J. P. Hasseman, G. Tsegaye, A. Tsang, A. Wong, R. Patel, J. J. Macklin, Y. Chen, A. Konnerth, V. Jayaraman, L. L. Looger, E. R. Schreiter, K. Svoboda, and D. S. Kim, "High-performance calcium sensors for imaging activity in neuronal populations and microcompartments," Nature Methods **16**, 649–657 (2019).

8. J. Huisken, J. Swoger, F. D. Bene, J. Wittbrodt, and E. H. K. Stelzer, "Optical Sectioning Deep Inside Live Embryos by Selective Plane Illumination Microscopy," Science **305**, 1007–1009 (2004).
9. M. B. Ahrens, M. B. Orger, D. N. Robson, J. M. Li, and P. J. Keller, "Whole-brain functional imaging at cellular resolution using light-sheet microscopy," Nature Methods **10**, 413–420 (2013).
10. T. Panier, S. Romano, R. Olive, T. Pietri, G. Sumbre, R. Candelier, and G. Debrégeas, "Fast functional imaging of multiple brain regions in intact zebrafish larvae using Selective Plane Illumination Microscopy," Front. Neural Circuits **7**, (2013).
11. T. W. Dunn, Y. Mu, S. Narayan, O. Randlett, E. A. Naumann, C.-T. Yang, A. F. Schier, J. Freeman, F. Engert, and M. B. Ahrens, "Brain-wide mapping of neural activity controlling zebrafish exploratory locomotion," eLife **5**, e12741 (2016).
12. G. Migault, T. L. van der Plas, H. Trentesaux, T. Panier, R. Candelier, R. Proville, B. Englitz, G. Debrégeas, and V. Bormuth, "Whole-Brain Calcium Imaging during Physiological Vestibular Stimulation in Larval Zebrafish," Current Biology **28**, 3723-3735.e6 (2018).
13. N. Vladimirov, C. Wang, B. Höckendorf, A. Pujala, M. Tanimoto, Y. Mu, C.-T. Yang, J. D. Wittenbach, J. Freeman, S. Preibisch, M. Koyama, P. J. Keller, and M. B. Ahrens, "Brain-wide circuit interrogation at the cellular level guided by online analysis of neuronal function," Nature Methods **15**, 1117–1125 (2018).
14. G. Vanwalleghem, K. Schuster, M. A. Taylor, I. A. Favre-Bulle, and E. K. Scott, "Brain-Wide Mapping of Water Flow Perception in Zebrafish," J. Neurosci. **40**, 4130–4144 (2020).
15. N. Vladimirov, Y. Mu, T. Kawashima, D. V. Bennett, C.-T. Yang, L. L. Looger, P. J. Keller, J. Freeman, and M. B. Ahrens, "Light-sheet functional imaging in fictively behaving zebrafish," Nature Methods **11**, 883–884 (2014).
16. S. Wolf, W. Supatto, G. Debrégeas, P. Mahou, S. G. Kruglik, J.-M. Sintes, E. Beaurepaire, and R. Candelier, "Whole-brain functional imaging with two-photon light-sheet microscopy," Nat Methods **12**, 379–380 (2015).
17. S. Wolf, A. M. Dubreuil, T. Bertoni, U. L. Böhm, V. Bormuth, R. Candelier, S. Karpenko, D. G. C. Hildebrand, I. H. Bianco, R. Monasson, and G. Debrégeas, "Sensorimotor computation underlying phototaxis in zebrafish," Nat Commun **8**, 1–12 (2017).
18. M. Göppert-Mayer, "Über Elementarakte mit zwei Quantensprüngen," Annalen der Physik **401**, 273–294 (1931).
19. F. Emran, J. Rihel, A. R. Adolph, and J. E. Dowling, "Zebrafish larvae lose vision at night," PNAS **107**, 6034–6039 (2010).
20. Z. Lavagnino, F. C. Zanacchi, E. Ronzitti, and A. Diaspro, "Two-photon excitation selective plane illumination microscopy (2PE-SPIM) of highly scattering samples: characterization and application," Opt. Express **21**, 5998–6008 (2013).
21. J. Palero, S. I. Santos, D. Artigas, and P. Loza-Alvarez, "A simple scanless two-photon fluorescence microscope using selective plane illumination," Opt. express **18**, 8491–8498 (2010).
22. O. E. Olarte, J. Licea-Rodriguez, J. A. Palero, E. J. Gualda, D. Artigas, J. Mayer, J. Swoger, J. Sharpe, I. Rocha-Mendoza, R. Rangel-Rojo, and P. Loza-Alvarez, "Image formation by linear and nonlinear digital scanned light-sheet fluorescence microscopy with Gaussian and Bessel beam profiles," Biomed. Opt. Express, BOE **3**, 1492–1505 (2012).
23. V. Maioli, A. Boniface, P. Mahou, J. F. Ortas, L. Abdeladim, E. Beaurepaire, and W. Supatto, "Fast in vivo multiphoton light-sheet microscopy with optimal pulse frequency," Biomed. Opt. Express, BOE **11**, 6012–6026 (2020).
24. T. V. Truong, W. Supatto, D. S. Koos, J. M. Choi, and S. E. Fraser, "Deep and fast live imaging with two-photon scanned light-sheet microscopy," Nat Methods **8**, 757–760 (2011).
25. A. Maruyama, Y. Oshima, H. Kajiura-Kobayashi, S. Nonaka, T. Imamura, and K. Naruse, "Wide field intravital imaging by two-photon-excitation digital-scanned light-sheet microscopy (2p-DSLM) with a high-pulse energy laser," Biomed. Opt. Express, BOE **5**, 3311–3325 (2014).
26. M. Zhao, H. Zhang, Y. Li, A. Ashok, R. Liang, W. Zhou, and L. Peng, "Cellular imaging of deep organ using two-photon Bessel light-sheet nonlinear structured illumination microscopy," Biomed. Opt. Express, BOE **5**, 1296–1308 (2014).
27. T. V. Truong, D. B. Holland, S. Madaan, A. Andreev, K. Keomanee-Dizon, J. V. Troll, D. E. S. Koo, M. J. McFall-Ngai, and S. E. Fraser, "High-contrast, synchronous volumetric imaging with selective volume illumination microscopy," Commun Biol **3**, 1–8 (2020).
28. K. Keomanee-Dizon, S. E. Fraser, and T. V. Truong, "A versatile, multi-laser twin-microscope system for light-sheet imaging," Review of Scientific Instruments **91**, 053703 (2020).
29. G. de Vito, L. Turrini, C. Fornetto, P. Ricci, C. Müllenbroich, G. Sancataldo, E. Trabalzini, G. Mazzamuto, N. Tiso, L. Sacconi, D. Fanelli, L. Silvestri, F. Vanzi, and F. S. Pavone, "Two-photon light-sheet microscopy for high-speed whole-brain functional imaging of zebrafish neuronal physiology and pathology," in *Neurophotonics* (International Society for Optics and Photonics, 2020), Vol. 11360, p. 1136004.
30. G. de Vito, C. Fornetto, P. Ricci, C. Müllenbroich, G. Sancataldo, L. Turrini, G. Mazzamuto, N. Tiso, L. Sacconi, D. Fanelli, L. Silvestri, F. Vanzi, and F. S. Pavone, "Two-photon high-speed light-sheet volumetric imaging of brain activity during sleep in zebrafish larvae," in *Neural Imaging and Sensing 2020* (International Society for Optics and Photonics, 2020), Vol. 11226, p. 1122604.
31. K. Podgorski and G. Ranganathan, "Brain heating induced by near-infrared lasers during multiphoton microscopy," Journal of Neurophysiology **116**, 1012–1023 (2016).

32. S. C. Baraban, M. R. Taylor, P. A. Castro, and H. Baier, "Pentylenetetrazole induced changes in zebrafish behavior, neural activity and c-fos expression," Neuroscience **131**, 759–768 (2005).
33. T. Afrikanova, A.-S. K. Serruys, O. E. M. Buenafe, R. Clinckers, I. Smolders, P. A. M. de Witte, A. D. Crawford, and C. V. Esguerra, "Validation of the Zebrafish Pentylenetetrazol Seizure Model: Locomotor versus Electrographic Responses to Antiepileptic Drugs," PLOS ONE **8**, e54166 (2013).
34. A. M. Stewart, D. Desmond, E. Kyzar, S. Gaikwad, A. Roth, R. Riehl, C. Collins, L. Monnig, J. Green, and A. V. Kalueff, "Perspectives of zebrafish models of epilepsy: What, how and where next?," Brain Research Bulletin **87**, 135–143 (2012).
35. L. Kou, D. Labrie, and P. Chylek, "Refractive indices of water and ice in the 0.65- to 2.5-μm spectral range," Appl. Opt., AO **32**, 3531–3540 (1993).
36. M. C. Müllenbroich, L. Turrini, L. Silvestri, T. Alterini, A. Gheisari, N. Tiso, F. Vanzi, L. Sacconi, and F. S. Pavone, "Bessel Beam Illumination Reduces Random and Systematic Errors in Quantitative Functional Studies Using Light-Sheet Microscopy," Front. Cell. Neurosci. **12**, (2018).
37. "ZFIN Feature: b4," https://zfin.org/ZDB-ALT-980203-365.
38. L. Turrini, C. Fornetto, G. Marchetto, M. C. Müllenbroich, N. Tiso, A. Vettori, F. Resta, A. Masi, G. Mannaioni, F. S. Pavone, and F. Vanzi, "Optical mapping of neuronal activity during seizures in zebrafish," Sci Rep **7**, 1–12 (2017).
39. M. Westerfield, *The Zebrafish Book: A Guide for the Laboratory Use of Zebrafish (Danio Rerio)*, 4th ed. (Univ. of Oregon Press, 2000).
40. C. K. Knox, "Detection of neuronal interactions using correlation analysis," Trends in Neurosciences **4**, 222–225 (1981).
41. A. Gałecki and T. Burzykowski, *Linear Mixed-Effects Models Using R: A Step-by-Step Approach*, Springer Texts in Statistics (Springer-Verlag, 2013).
42. R Core Team, *R: A Language and Environment for Statistical Computing* (R Foundation for Statistical Computing, 2015).
43. A. Kuznetsova, P. B. Brockhoff, and R. H. B. Christensen, "lmerTest Package: Tests in Linear Mixed Effects Models," J STAT SOFTW **82**, (2017).
44. P.-C. Bürkner, "brms: An R Package for Bayesian Multilevel Models Using Stan," Journal of Statistical Software **80**, 1–28 (2017).
45. G. Sancataldo, V. Gavryusev, G. de Vito, L. Turrini, M. Locatelli, C. Fornetto, N. Tiso, F. Vanzi, L. Silvestri, and F. S. Pavone, "Flexible Multi-Beam Light-Sheet Fluorescence Microscope for Live Imaging Without Striping Artifacts," Front. Neuroanat. **13**, (2019).
46. P. Ricci, V. Gavryusev, C. Müllenbroich, L. Turrini, G. de Vito, L. Silvestri, G. Sancataldo, and F. S. Pavone, "Removing striping artifacts in light-sheet fluorescence microscopy: a review," Progress in Biophysics and Molecular Biology, (posted 15 July 2021, in press).
47. G. de Vito, P. Ricci, L. Turrini, V. Gavryusev, M.-C. Muellenbroich, N. Tiso, F. Vanzi, L. Silvestri, and F. S. Pavone, "Effects of excitation light polarization on fluorescence emission in two-photon light-sheet microscopy," Biomedical Optics Express **11**, 4651–4665 (2020).
48. F. O. Fahrbach, F. F. Voigt, B. Schmid, F. Helmchen, and J. Huisken, "Rapid 3D light-sheet microscopy with a tunable lens," Opt. Express, OE **21**, 21010–21026 (2013).
49. J. Ono, R. F. Vieth, and P. D. Walson, "Electrocorticographical observation of seizures induced by pentylenetetrazol (PTZ) injection in rats," Funct Neurol **5**, 345–352 (1990).
50. M. Kunst, E. Laurell, N. Mokayes, A. Kramer, F. Kubo, A. M. Fernandes, D. Förster, M. D. Maschio, and H. Baier, "A Cellular-Resolution Atlas of the Larval Zebrafish Brain," Neuron **103**, 21-38.e5 (2019).
51. O. Randlett, C. L. Wee, E. A. Naumann, O. Nnaemeka, D. Schoppik, J. E. Fitzgerald, R. Portugues, A. M. B. Lacoste, C. Riegler, F. Engert, and A. F. Schier, "Whole-brain activity mapping onto a zebrafish brain atlas," Nature Methods **12**, 1039–1046 (2015).
52. R. S. Fisher, H. E. Scharfman, and M. deCurtis, "How can we identify ictal and interictal abnormal activity?," Adv Exp Med Biol **813**, 3–23 (2014).
53. C. Diaz Verdugo, S. Myren-Svelstad, E. Aydin, E. Van Hoeymissen, C. Deneubourg, S. Vanderhaeghe, J. Vancraeynest, R. Pelgrims, M. I. Cosacak, A. Muto, C. Kizil, K. Kawakami, N. Jurisch-Yaksi, and E. Yaksi, "Glia-neuron interactions underlie state transitions to generalized seizures," Nature Communications **10**, 3830 (2019).
54. J. Liu and S. C. Baraban, "Network Properties Revealed during Multi-Scale Calcium Imaging of Seizure Activity in Zebrafish," eNeuro **6**, (2019).
55. M. J. Winter, D. Windell, J. Metz, P. Matthews, J. Pinion, J. T. Brown, M. J. Hetheridge, J. S. Ball, S. F. Owen, W. S. Redfern, J. Moger, A. D. Randall, and C. R. Tyler, "4-dimensional functional profiling in the convulsant-treated larval zebrafish brain," Scientific Reports **7**, 6581 (2017).
56. R. E. Rosch, P. R. Hunter, T. Baldeweg, K. J. Friston, and M. P. Meyer, "Calcium imaging and dynamic causal modelling reveal brain-wide changes in effective connectivity and synaptic dynamics during epileptic seizures," PLOS Computational Biology **14**, e1006375 (2018).
57. M. J. Winter, J. Pinion, A. Tochwin, A. Takesono, J. S. Ball, P. Grabowski, J. Metz, M. Trznadel, K. Tse, W. S. Redfern, M. J. Hetheridge, M. Goodfellow, A. D. Randall, and C. R. Tyler, "Functional brain imaging in larval zebrafish for characterising the effects of seizurogenic compounds acting via a range of pharmacological mechanisms," British Journal of Pharmacology **178**, 2671–2689 (2021).

58. O. Cozzolino, F. Sicca, E. Paoli, F. Trovato, F. M. Santorelli, G. M. Ratto, and M. Marchese, "Evolution of Epileptiform Activity in Zebrafish by Statistical-Based Integration of Electrophysiology and 2-Photon $Ca^{2+}$ Imaging," Cells **9**, 769 (2020).
59. S. Hong, P. Lee, S. C. Baraban, and L. P. Lee, "A Novel Long-term, Multi-Channel and Non-invasive Electrophysiology Platform for Zebrafish," Sci Rep **6**, 28248 (2016).
60. L. Chicchi, G. Cecchini, I. Adam, G. de Vito, R. Livi, F. S. Pavone, L. Silvestri, L. Turrini, F. Vanzi, and D. Fanelli, "Reconstruction scheme for excitatory and inhibitory dynamics with quenched disorder: application to zebrafish imaging," J Comput Neurosci **49**, 159–174 (2021).

# Fast whole-brain imaging of seizures in zebrafish larvae by two-photon light-sheet microscopy: supplemental document

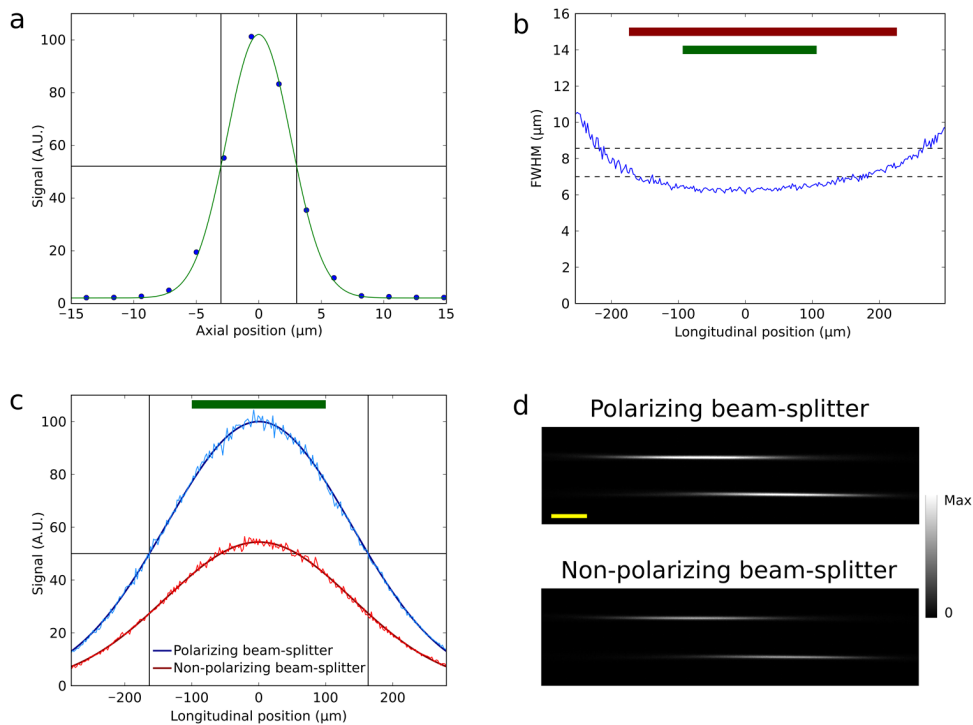

Fig. S1. (a) Transversal profile of the excitation beam at the waist. Gaussian fit in green. The single horizontal and the two vertical lines indicate the half-maximum value and the FWHM range, respectively. (b) Graph of the transversal FWHM values of the excitation beam depending on the longitudinal position along the beam. The zero value on the x-axis denotes the waist. The two dashed horizontal lines indicate the FWHM at the waist increased by a factor of $\sqrt{2}$ (8.56 μm, used to identify the confocal parameter) and the value 7 μm. The red and the green bars on the top indicate the typical size of the larval brain along the lateral dimension and half this value, respectively. (c) Longitudinal profiles of the excitation beam generated employing a polarizing (blue) or a non-polarizing (red) beam-splitters. Gaussian fitting lines (data) in darker (lighter) colors. The single horizontal and the two vertical lines indicate the half-maximum value and the FWHM range, respectively, referring to the polarizing beam-splitter condition. The green bar on the top indicates half the typical size of the larval brain along the lateral dimension. (d) Images of the two excitation beams in fluorescein solution generated employing a polarizing (top) or a non-polarizing (bottom) beam-splitters. The two beams were displaced relatively to each other along the transversal dimension purposely to improve their visualization. Scale bar: 100 μm.

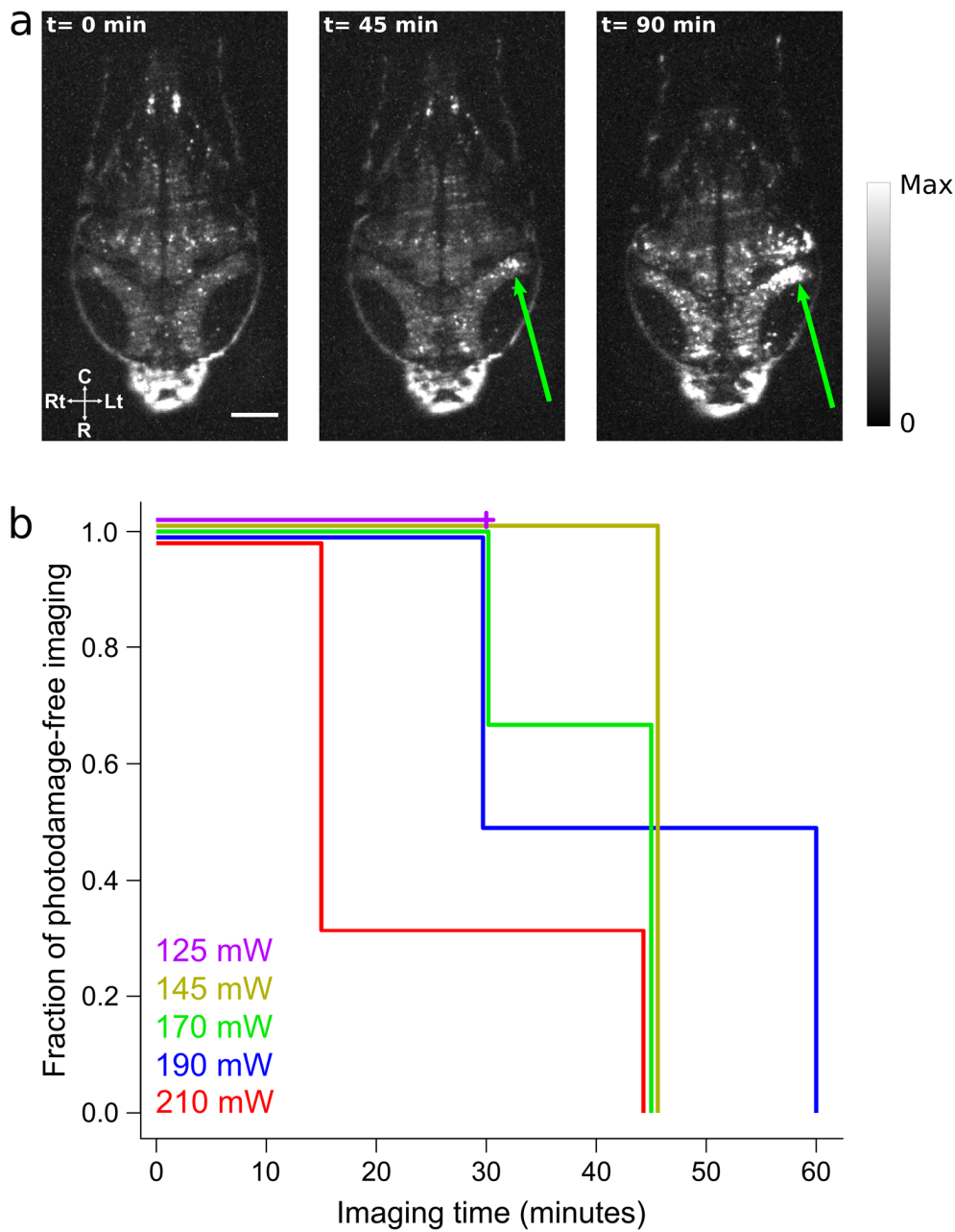

Fig. S2. (a) Coronal sections of the larval brain before the (left), at the (center) and after the (right) insurgence of the photodamage, defined as the appearance of an area of tonic neuronal calcium activity (indicated by the green arrows). The uninterrupted imaging time is reported on the panels. The images are temporal averages over 2 seconds. R: rostral, C: caudal, Rt: right, Lt: left. Scale bar: 100 μm. (b) Survival plot depicting the fraction of larvae not showing photodamage for each time-point and tested power. Right-censoring in the 125-mW data is indicated by a cross. Adjacent lines were slightly shifted to improve visualization. Different colors indicate different cumulative excitation power levels measured at the exit of the objectives, as specified in the legend; the excitation power at the sample is estimated to be reduced by 33%. Raw data shown in Table S1.

**Table S1. Cumulative illumination power at the exit of the objectives and time of appearance of photodamage for each tested larva.**

| Power (mW) | Time to photodamage (minutes) |
|---|---|
| 125 | >30 |
| 145 | 45 |
| 170 | 45 |
| 170 | 45 |
| 170 | 30 |
| 190 | 30 |
| 190 | 60 |
| 210 | 45 |
| 210 | 15 |
| 210 | 15 |

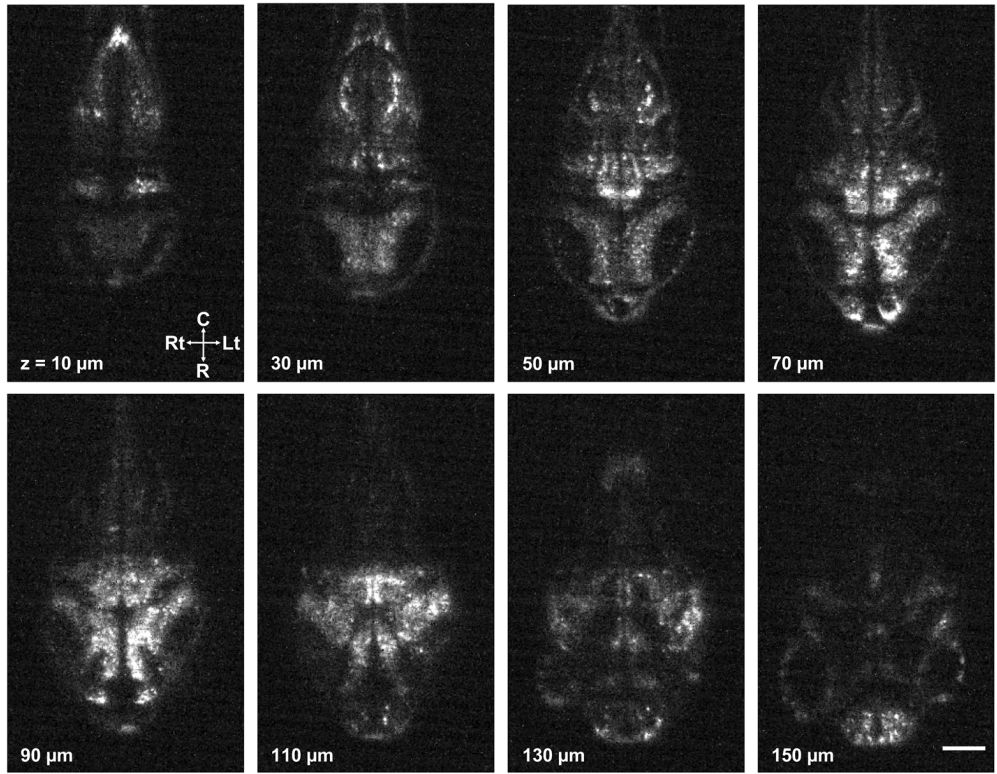

Fig. S3. Eight coronal sections showing a single temporal frame from the same time-lapse volumetric recording of Fig. 2b. The sections were recorded at different dorso-ventral depths of the larva (indicated on the panels, with respect to the dorsal surface). R: rostral, C: caudal, Rt: right, Lt: left. Scale bar: 100 μm.

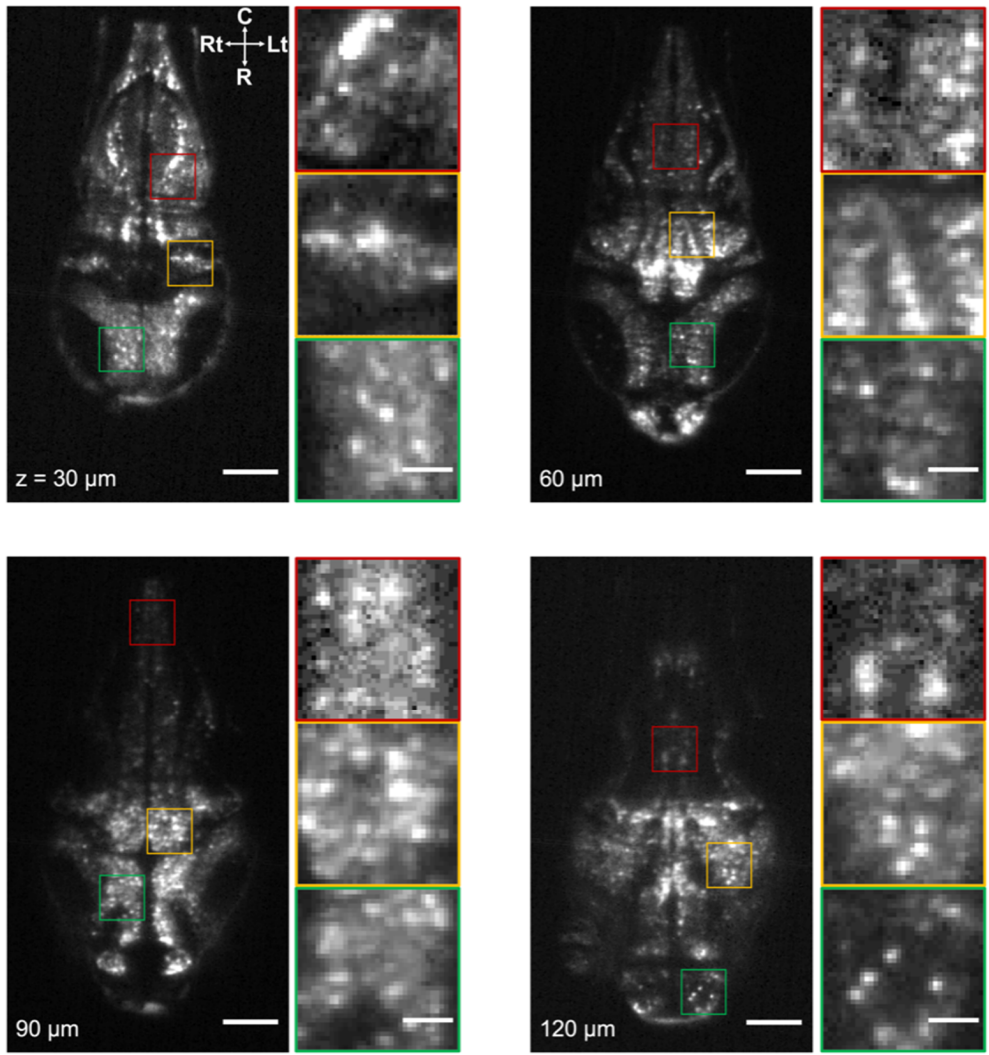

Fig. S4. Main panels: four coronal sections showing temporal average intensity projections at different dorso-ventral depths of a larval brain (indicated on the panels, with respect to the dorsal surface). R: rostral, C: caudal, Rt: right, Lt: left. Scale bar: 100 μm. Side panels: magnifications of the areas indicated by the rectangles of the same respective colors on the main panels. Brightness and contrast were optimized separately for each image to improve visibility. Scale bar: 25 μm.

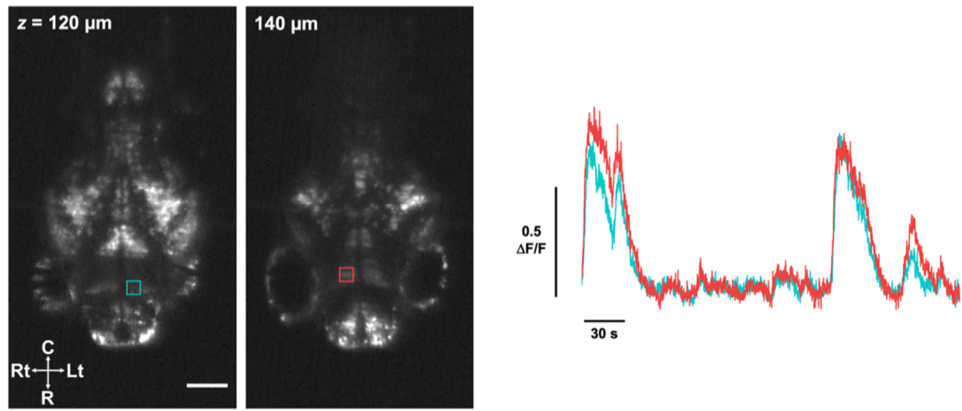

Fig. S5. Left: two coronal sections showing temporal average intensity projections at different dorso-ventral depths of a larval brain (indicated on the panels, with respect to the dorsal surface) affected by reduced image quality in the forebrain area in between the eyes. R: rostral, C: caudal, Rt: right, Lt: left. Scale bar: 100 μm. Right: plot of neuronal activity over time extracted from the forebrain ROIs drawn on left panels.

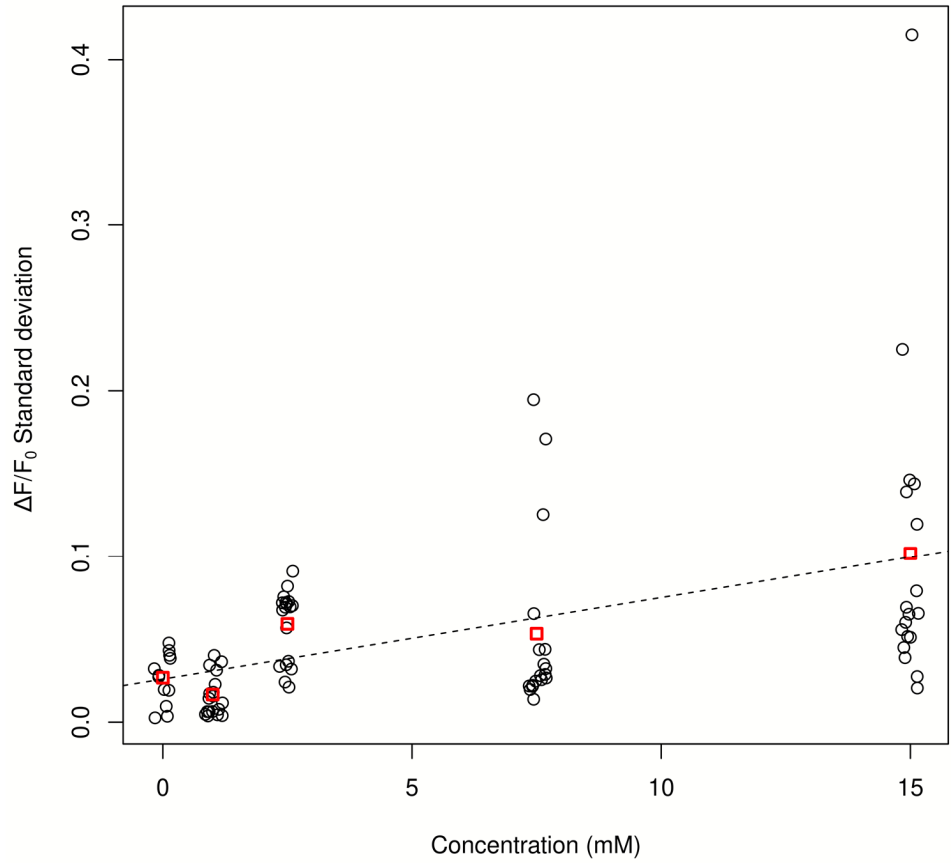

Fig. S6. Scatter plot of the standard deviation values of whole-brain-averaged $\Delta F/F_0$ traces (black hollow circles) against the PTZ concentrations. A small amount of jitter was applied on the x-axis to improve visualization. Red squares denote the average values for each PTZ concentration. The dashed line indicates the linear regression (p-value < 0.001, $R^2 \approx 0.22$).(1)

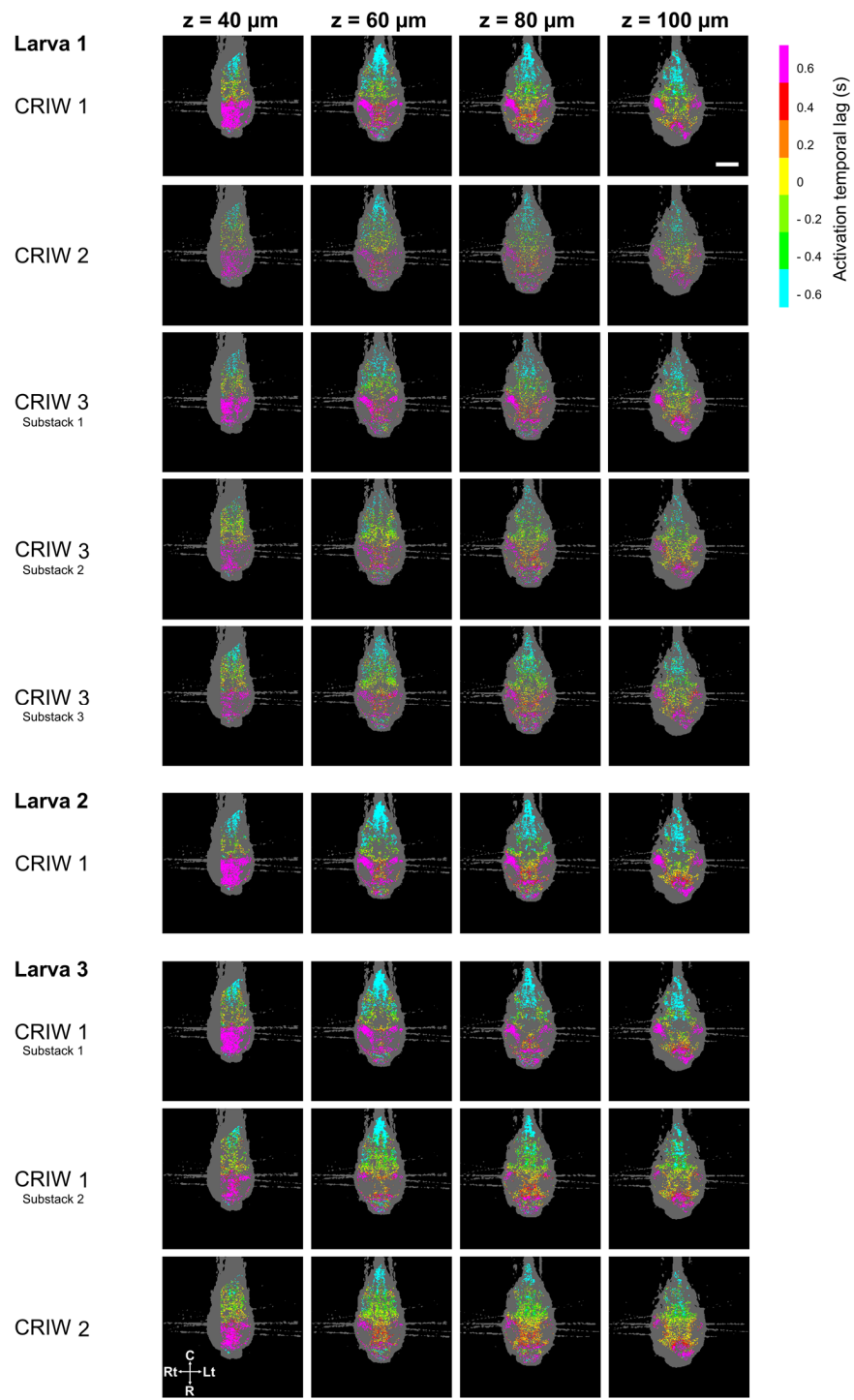

Fig. S7. These color-maps display, for each recorded CRIW event (shown along rows) from the different larvae, the voxels having the same lag-value (color) as in the aggregated color-map shown in Fig. 4c, central panel ("ICTAL"). Coronal sections at different dorso-ventral depths of a larva (indicated on the top, with respect to the dorsal surface) are shown across columns. Unmasked voxels with non-concordant lag values are displayed in grey. Color code as in Fig. 4a,c,e and as specified in the color-bar on the top-right. R: rostral, C: caudal, Rt: right, Lt: left. Scale bar: 100 μm.

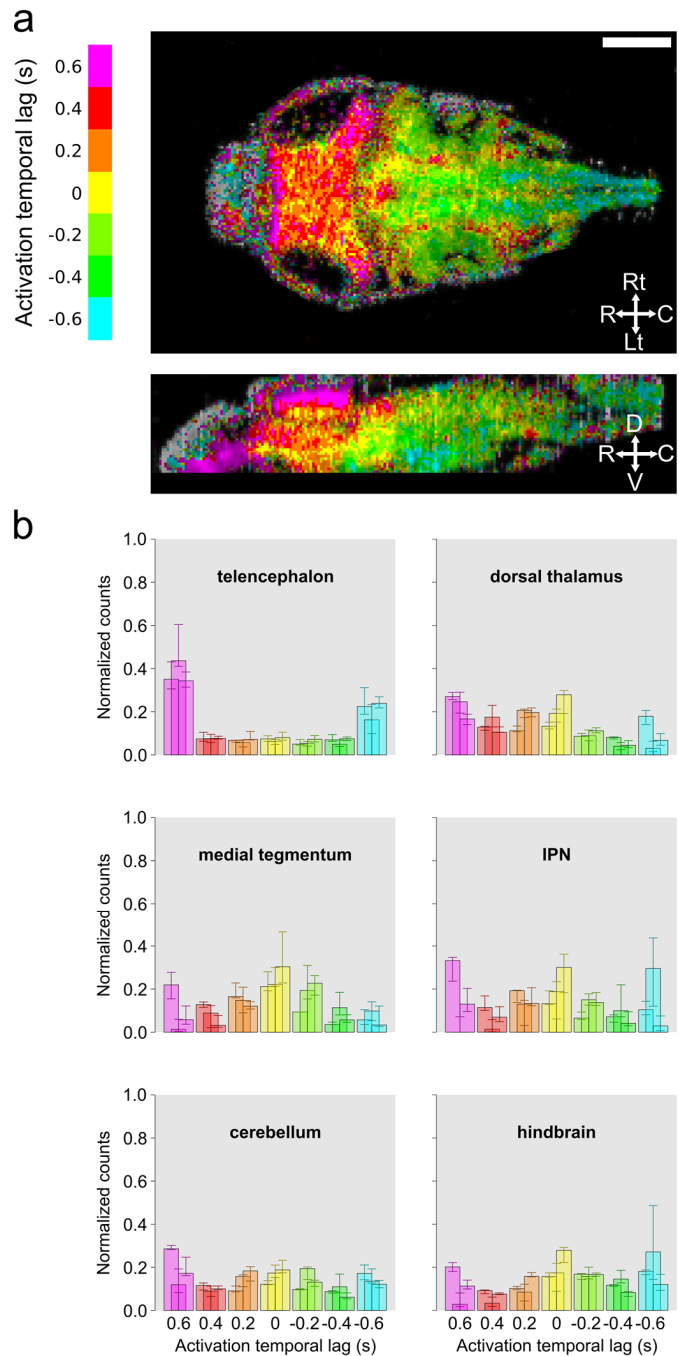

Fig. S8. (a) Lag map for a single ictal event corresponding to the calcium traces shown in Fig. 4b. Top: coronal section; bottom: sagittal section. The voxel hues encode the lag value as specified by the color map on the left and as in Fig. 4a,c,e. Masked or non-statistically significant voxels are depicted in grey values. Voxel brightness is a function of its time-averaged $\Delta F/F_0$ value. R: rostral, C: caudal, D: dorsal, V: ventral, Rt: right, Lt: left. Scale bar: 100 μm. (b) Bar plots showing the relative frequencies of the voxel-based lag values for each ROI not shown in Fig. 4e (as indicated on the plots). The bar colors correspond to the color code used in (a). The height of each bar corresponds to the median value of the single-acquisition-based frequency distribution, while the error bars indicate the first and the third quartiles of the distribution. The first, second and third bars in each lag class correspond to the control, ictal and postictal situations, respectively.

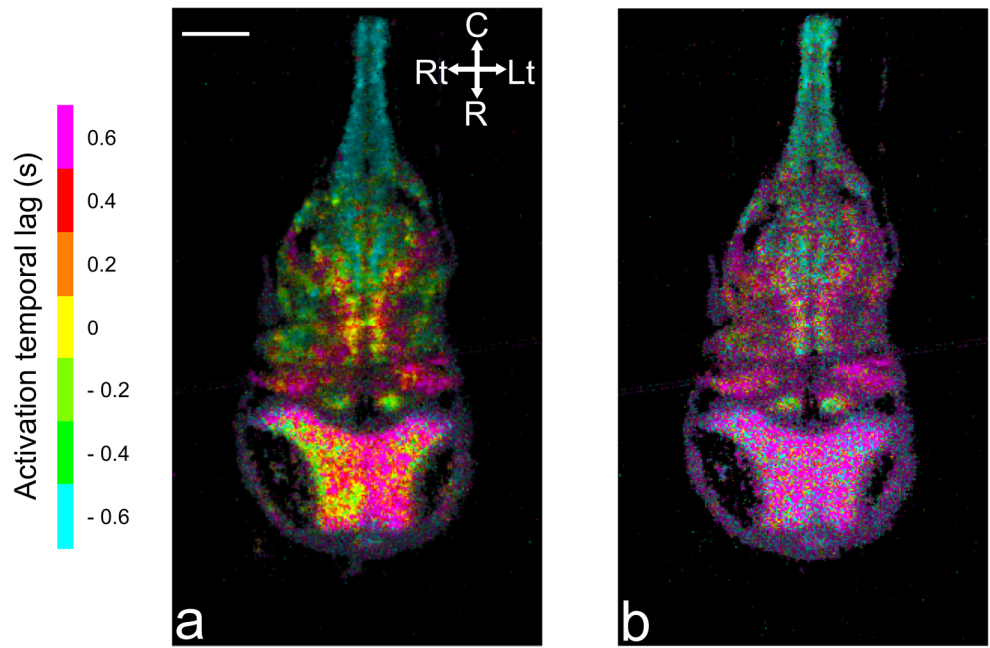

Fig. S9. Lag maps (coronal sections) for a single ictal event. (a) Results obtained with the cross-correlation based method (described in section "*2.8 Voxel-based lag analysis*"). R: rostral, C: caudal, Rt: right, Lt: left. (b) Results obtained with the time-derivative-method described in [54]. The voxel hues encode the lag value as specified by the color map on the left and as in Fig. 4a,c,e. The voxel saturation is kept at maximum. Scale bar: 100 μm.

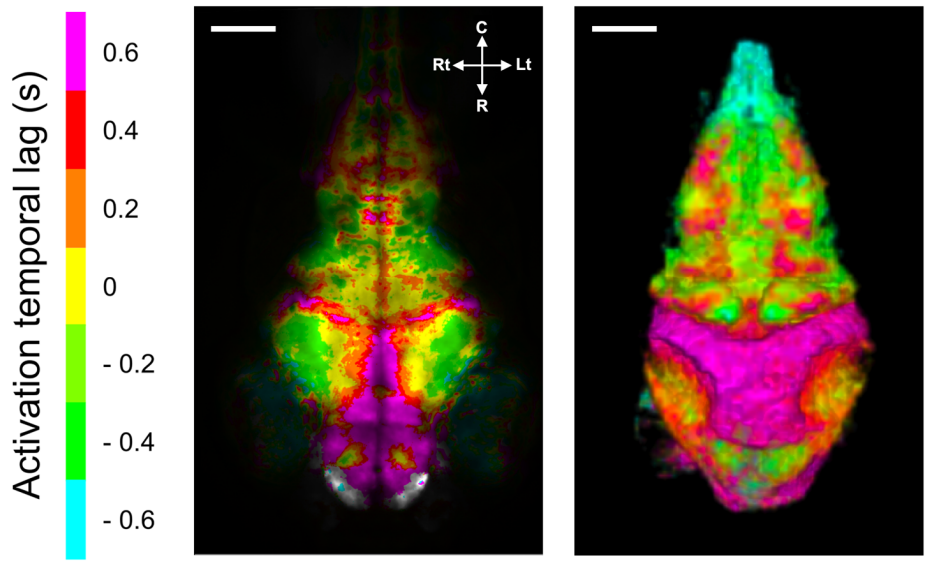

Fig. S10. Median lag map of three ictal events recorded using 1P wide-field fluorescence microscopy in a 4-dpf Tg(elavl3:GCaMP6s) zebrafish larva treated with 15 mM PTZ (left) and dorsal view of the lag map during ictal activity presented in Fig. 4a (right). The voxel hues encode the lag value as specified by the color map on the left and as in Fig. 4a,c,e. Non-statistically significant voxels are depicted in grey values. R: rostral, C: caudal, Rt: right, Lt: left. Scale bar: 100 μm.

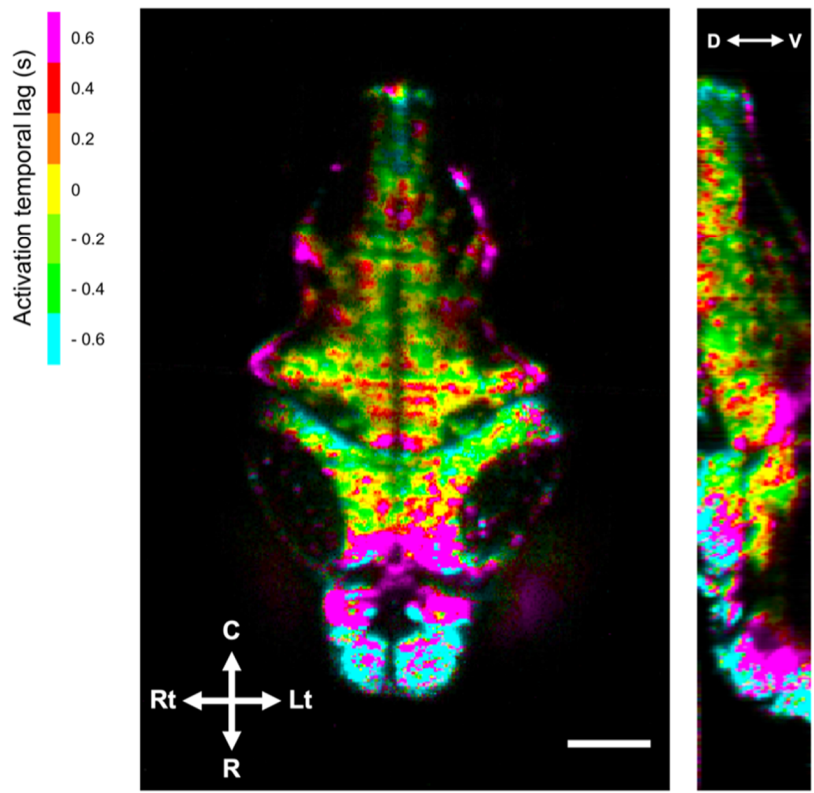

Fig. S11. Median lag map (left: coronal section; right: sagittal section) of 6 ictal events recorded in 6 dpf Tg(H2B-GCaMP6s) larvae exposed to 10 mM PTZ, as in Ref. [54]. The voxel hues encode the lag value as specified by the color map on the left and as in Fig. 4a,c,e. R: rostral, C: caudal, D: dorsal, V: ventral, Rt: right, Lt: left. Scale bar: 100 μm.